\newif\ifonecol 
\onecolfalse 

\ifonecol
\documentclass[journal,12pt,onecolumn]{IEEEtran}
\else
\documentclass[journal]{IEEEtran}
\fi

\newlength{\figurewidth}
\ifonecol
\else
\fi

\ifonecol
\usepackage{setspace}
\fi

\usepackage{graphicx}
\usepackage{afterpage}
\usepackage{amsmath}
\usepackage{amsfonts}
\usepackage{amsbsy}
\usepackage{amssymb}
\usepackage[noadjust]{cite}
\usepackage{bm}
\usepackage{epsf}
\usepackage[all]{xy}
\usepackage{epsfig}
\usepackage{subfigure}
\usepackage{stfloats}
\usepackage{floatrow}

\newtheorem{theorem}{Theorem}

\newcommand{\pf}{\noindent{\bf Proof:~}}
\newcommand{\qedsymb}{\hfill{\rule{2mm}{2mm}}}
\DeclareMathAlphabet{\mathsfbf}{OT1}{cmss}{sbc}{n}

\newcommand{\EE}{\mathbb{E}} 

\newcommand{\xv}{{\bf x}}
\newcommand{\yv}{{\bf y}}

\newcommand{\Fm}{{\bf F}}

\newcommand{\Hm}{{\bf H}}
\newcommand{\Id}{{\bf I}}
\newcommand{\Jm}{{\bf J}}

\newcommand{\Mm}{{\bf M}}

\newcommand{\Xm}{{\bf X}}
\newcommand{\Ym}{{\bf Y}}
\newcommand{\Zm}{{\bf Z}}

\newcommand{\phiv}{\boldsymbol{\phi}}

\newcommand{\Gammam}{\boldsymbol{\Gamma}}

\newcommand{\Sigmam}{\boldsymbol{\Sigma}}
\newcommand{\Phim}{\boldsymbol{\Phi}}

\newcommand{\Omegam}{\boldsymbol{\Omega}}

\def\Tran{\mathsf{^T}}

\def\ben{\begin{enumerate}}
\def\beq{\begin{equation}}
\def\beqa{\begin{eqnarray}}
\def\bit{\begin{itemize}}
\def\een{\end{enumerate}}
\def\eeq{\end{equation}}
\def\eeqa{\end{eqnarray}}
\def\eit{\end{itemize}}

\def\non{\nonumber\\}

\newtheorem{proposition}[theorem]{Proposition}

\ifCLASSINFOpdf
\else
\fi

\newif\ifforarxiv
\forarxivtrue

\ifforarxiv
\newcommand{\MYfooter}{\smash{\scriptsize
\hfil\parbox[t][\height][t]{\textwidth}{\centering
This work has been accepted for publication at Journal of Communications and Networks.}\hfil\hbox{}}}

\makeatletter

\def\ps@IEEEtitlepagestyle{%
\def\@oddfoot{\MYfooter}%
\def\@evenfoot{\MYfooter}}

\makeatother

\fi

\begin{document}
%
\title{Compensation of Phase Noise in Massive-MIMO Uplink Communications Based on Expectation-Maximization Algorithm}
%
%
%

\author{Alberto Tarable and Francisco J. Escribano,~\IEEEmembership{Senior Member,~IEEE}%
\thanks{Alberto Tarable is with the Consiglio Nazionale delle Ricerche, Istituto di Elettronica e di Ingegneria Informatica e delle Telecomunicazioni (CNR-IEIIT), Italy (e-mail: alberto.tarable@ieiit.cnr.it).}%
\thanks{ Francisco J. Escribano is with the Department of Signal Theory and Communications, Universidad de Alcal\'a, 28805 Alcal\'a de Henares, Spain (e-mail: francisco.escribano@uah.es).}}

\maketitle

\begin{abstract}
Phase noise (PN) is a major disturbance in MIMO systems, where the contribution of different oscillators at the transmitter and the receiver side may degrade the overall performance and offset the gains offered by MIMO techniques. This is even more crucial in the case of massive MIMO, since the number of PN sources may increase considerably. In this work, we propose an iterative receiver based on the application of the expectation-maximization algorithm. We consider a massive MIMO framework with a general association of oscillators to antennas, and include other channel disturbances like imperfect channel state information and Rician block fading. At each receiver iteration, given the information on the transmitted symbols, steepest descent is used to estimate the PN samples, with an optimized adaptive step size and a threshold-based stopping rule. The results obtained for several test cases show how the bit error rate and mean square error can benefit from the proposed phase-detection algorithm, even to the point of reaching the same performance as in the case where no PN is present, offering better results than a state-of-the-art alternative. Further analysis of the results allow to draw some useful trade-offs respecting final performance and consumption of resources.
\end{abstract}

\begin{IEEEkeywords}
Massive MIMO, EM Algorithm, Phase Noise, Iterative Decoding
\end{IEEEkeywords}

%

\section{Introduction}

Massive MIMO \cite{Marzetta10} is a key enabling technique to achieve the large throughput required by current 5G cellular networks. In the massive MIMO concept, the base station (BS) is equipped with many more antennas than the number strictly needed to communicate with the users in its cell. Thanks to this overabundance of antennas, users communicating in the same time-frequency resources can be distinguished by means of simple linear filter techniques, such as minimum mean-square error (MMSE) in the uplink, or maximum ratio transmission (MRT) \cite{Lo99} in the downlink. Such filters require channel state information (CSI) for all links, in order to behave properly. CSI is typically obtained through the transmission of pilots, which are inserted within the frame at regular time intervals, depending on the channel coherence time. Thus, in order to achieve the performance promised by massive MIMO, obtaining a good CSI estimation becomes of crucial importance.

In real systems, the channel seen at the receiver is time-varying not only because of the presence of mobility, either of the user equipments (UEs) or of the surrounding environment (e.g., scatterers, reflectors, etc.), but also because of hardware nonidealities. One of the most important impairments of this type is phase noise (PN) \cite{Bjornson14}. PN arises from instability in the oscillators and typically assumes the form of a multiplicative noise, whose effect in massive MIMO systems can become particularly severe, substantially harming the performance of the receiver filters. While, in principle, pilot transmission allows the estimation of actual PN, these processes usually have a much faster time evolution than other forms of channel variations. Thus, between two pilot transmissions, the estimated PN becomes rapidly obsolete, i.e. information aging takes place.

In this paper, we deal with the problem of reducing the impact of PN in a single-cell massive MIMO uplink.
In particular, we introduce a receiver that is able to approximate joint PN detection and demodulation by iterating between a phase detector and a demodulator/decoder. The entire receiver is based on the expectation-maximization (EM) algorithm. The receiver was originally introduced in \cite{Tarable13} for a non-massive MIMO line-of-sight (LoS) system, where the proposed structure was shown to achieve excellent performance.

\subsection{Related work}

Several papers from the literature have faced the problem of PN in massive MIMO systems. A major distinction can be made between papers that deal with single-carrier transmission schemes and papers that deal with OFDM. 

For massive MIMO-OFDM systems, the problem of PN is even more relevant than in single-carrier systems, as PN causes the loss of orthogonality of the subcarriers, giving rise to inter-carrier interference (ICI). Apart from the large bulk of literature regarding the impact of PN on non-massive MIMO-OFDM systems, the work in \cite{Krishnan14} specifically considers the uplink of a massive MIMO-OFDM system affected by PN. Two extreme cases are considered, the synchronous case where a single oscillator is feeding all the $M$ BS antennas, and the asynchronous case where each BS antenna is fed by a different oscillator. The ergodic capacity in the asymptotic limit $M \rightarrow \infty$ is computed and a Kalman filter is proposed to track the PN. Under a similar scenario, in \cite{Pitarokoilis16} the authors derive lower bounds to the ergodic capacity for a finite value of $M$. In reference \cite{Puglielli16}, a more composite PN model is introduced, in which the phase-locked loop (PLL) structure is taken into account to derive the PN statistics. It is shown that, while independent PN at each BS antenna is averaged out, a common PN component can give a substantial performance loss. In \cite{Corvaja16}, the downlink of a massive MIMO-OFDM system is considered and the impact of PN on the performance of the zero-forcing and MRT precoders is characterized. The effect of imperfect CSI is also investigated. In \cite{Cheng19}, an iterative algorithm, based on variational EM, is proposed for compensation of PN in the uplink of massive MIMO-OFDM systems. Such algorithm performs joint channel and PN estimation. In a very recent paper (\cite{Qiao22}), an iterative PN and channel estimation algorithm for mmWave massive MIMO-OFDM systems is presented, where PN is estimated with pattern search and channel estimation is based on compressed sensing.

For single-carrier systems, the authors in \cite{Pitarokoilis15} consider the uplink of frequency-selective massive MIMO systems, with single-antenna users transmitting to the BS. Again, synchronous and asynchronous cases are considered for the PN at the BS.  A time-reversal maximal-ratio combining is proposed, and its performance is analyzed. Asymptotically in $M$, the array gain is shown to be $O (\sqrt{M})$, as in the case without PN. In the non-asymptotic case, a dramatic worsening of performance due to PN is shown, with a further loss in the asynchronous case. The performance loss is also caused by the very simple receiver considered. The downlink of a frequency-flat massive MIMO system is studied in \cite{Krishnan16}. In the downlink, it is unlikely to introduce a PN estimation/tracking algorithm, due to the limited computational complexity of the UEs. In \cite{Krishnan16}, the PN model at the BS is more general, supposing that an oscillator can feed several BS antennas. The effect of PN on the per-user SINR is evaluated, depending on the different linear precoder, i.e., zero-forcing (ZF), regularized ZF, or matched filter. It is shown that the performance loss increases with an increasing number of oscillators. In \cite{Wang18} a precoding scheme based on ZF with PN suppression is proposed for the downlink of a massive MIMO with time division duplexing (TDD). Due to the reciprocity property, the estimated uplink channel (obtained with a simple PN estimation scheme) is used in the downlink to form the ZF precoder. The work in \cite{Yang19} introduces an algorithm based on approximate Bayesian interference for symbol error reduction of the PN-impaired uplink of massive MIMO systems. 

More recently, reference \cite{Rasekh21} investigates PN-affected massive MIMO systems in the mmWave band. The PN model is similar to the one considered in \cite{Puglielli16}, with a common reference low-frequency oscillator that feeds all BS antennas, and a bank of PLLs, each feeding a subset of BS antennas, in order to raise the carrier frequency to the mmWave band. Thus, there are two contributions to PN, one from the reference oscillator, common to all antennas, the other from the  PLL voltage-controlled oscillator (VCO), which is independent for different antenna subsets. The first PN contribution is proven to be low-frequency and thus easily estimated, while the second term has a faster dynamics and thus turns out to be more problematic. The performance is characterized for MMSE filtering, when QPSK symbols are transmitted, yielding assessment of the PN impact on the output BER. In \cite{Corvaja20}, the focus is on hybrid analog-digital schemes for mmWave massive MIMO, for which the sensitivity to PN and channel estimation errors is quantitatively analyzed. Hybrid analog-digital schemes are a way to reduce the number of RF chains, which is especially important at the mmWave band. Moreover, a comparison between the hybrid and the fully-digital scheme in terms of robustness with respect to PN is performed. In \cite{Chatelier22} the impact of PN on beamforming performance for mmWave massive MIMO systems is investigated.

On the other hand, a very recent flavor that is capturing the attention of the research community is related to cell-free massive MIMO systems. The work in \cite{Fang21} analyzes the impact of PN on the performance of cell-free massive MIMO systems, both for their uplink and their downlink, showing that end users suffer more than access points (APs) from PN. PN-aware power control is also studied. In \cite{Jin21}, the spectral efficiency of a frequency-selective cell-free massive MIMO system with PN and imperfect channel estimation is studied, specifically when two linear low-complexity decoders, namely, time-reversal maximum-ratio combining and time-reversal large-scale fading decoding, are employed. Finally, it is worth noting that massive MIMO systems affected by PN have been studied in several particular applications, such as compressive channel estimation \cite{Zhang20}, PHY-layer authentication \cite{Zhang21}, or joint channel estimation and localization \cite{Zheng20}, among others.

In this paper, we consider the uplink of a single-carrier frequency-flat massive MIMO system. Thus, apart from the band, we are in a setup close to that of \cite{Rasekh21}. We consider independent PN processes for given antennas subsets, i.e., we neglect a common reference oscillator, whose PN contribution is proven in \cite{Rasekh21} to be easily estimated. The novelty of this work relies in a study of the performance improvement determined by an iterative receiver, which represents a viable, suboptimal implementation of a joint phase detector and demodulator/decoder. It is to be noted that some preliminary simulation results obtained with the receiver proposed in this paper were shown in \cite{Tarable21} to assess the effect of CSI aging. 

Summarizing, the main contributions of this paper are as follows:
\begin{itemize}

\item We describe in detail the proposed receiver under a general setting where each oscillator at the transmitter and at the receiver can feed an arbitrary number of antennas; 

\item We derive the Bayesian Cram\'er-Rao Bound (BCRB) for this general setting, yielding a benchmark for the performance in terms of the mean-square error (MSE) of the proposed receiver;

\item Through numerical simulations, we analyze the receiver performance both in terms of MSE and in terms of bit error rate (BER) in several realistic scenarios, including imperfect CSI;

\item We conclude with several design rules both for the receiver and for the system, in order to achieve a given target performance.
\end{itemize}

With respect to \cite{Tarable13}, \cite{Tarable21} and \cite{Tarable14}, the new contributions are:
\begin{itemize}
\item The model in \cite{Tarable13} has been generalized, since in this paper a given oscillator can feed more antennas; moreover, we consider here a massive MIMO system, possibly in non-LoS conditions, unlike \cite{Tarable13}, where the channel matrix was assumed to be unitary.

\item The step size of the steepest-descent algorithm used for phase detection has been optimized, whereas the architecture in \cite{Tarable13} used a fixed step size. 

\item The effect of fading and of imperfect channel state information is taken into account and investigated.

\item The performance of the proposed receiver is compared with that of the solution proposed in \cite{Yang19}, adapted to our scenario of coded transmission and Wiener PN model.

\item The Cram\'er-Rao bound is extended to the more general system model of this paper, while \cite{Tarable14} contains a derivation of the Cram\'er-Rao bound  for the scenario  in \cite{Tarable13}.

\item Many more simulation results are shown, in comparison with \cite{Tarable21}, which does not give the details of the algorithm. A thorough analysis of the algorithm complexity is given for the first time.
\end{itemize}

The remaining of this paper is structured as follows. In Section II, we describe the system under consideration, with Subsection II-B devoted to the description of the pilot transmission scheme and of the assumptions made on channel estimation. In Section III, the EM-based receiver is described, with particular emphasis on the phase estimator (Subsection III-A) and a brief description of the MIMO demodulator (Subsection III-B). Section IV derives the Bayesian Cram\'er-Rao bound. Section V shows the simulation results, both in terms of MSE (Subsection V-A), BER (Subsection V-B), and average number of receiver iterations and steepest-descent steps (Subsection V-C), and it looks as well into the algorithm complexity and latency (Subsection V-D). Finally, Section VI draws some conclusions.

\section{System description}

\subsection{Overall channel model}

The model we describe below, while general in nature, is specially suitable for the uplink of a wireless network in which $K$ users (typically, with one or few transmit antennas each) transmit to a base station equipped with many antennas and able to sustain a relevant computational burden.  We assume that there is a power-control mechanism so that all users are received with the same power. Besides being true for many real systems, such assumption constitutes a worst case for phase detection,  as unequal received power could arguably be of help in that respect. This situation will (implicitly or explicitly, where corresponding) constitute our specific context throughout the article.

\begin{figure}[!ht]
\centering
\centerline{\includegraphics[width=\figurewidth]{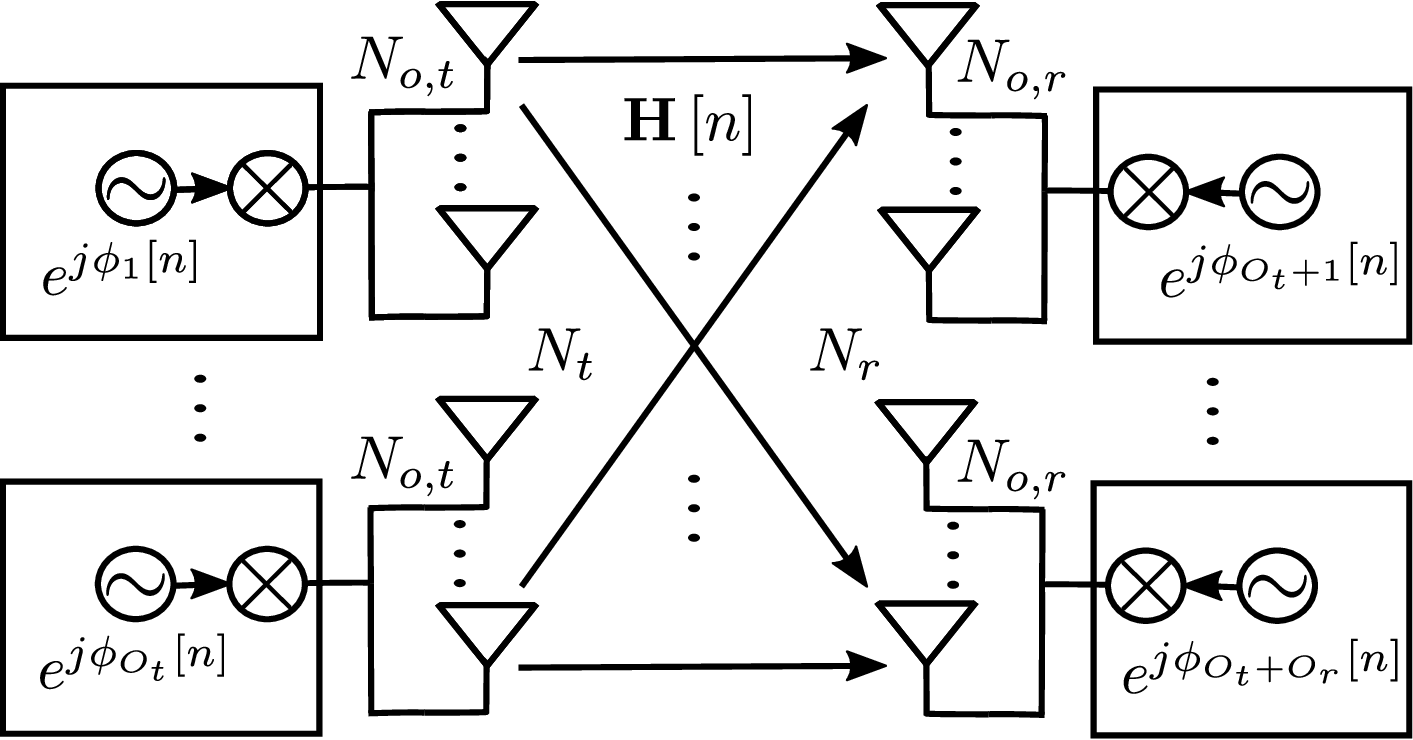}}
\caption{Overall scheme of the MIMO channel and oscillator/antenna setup for the massive MIMO system.}
\label{fig:System}
\end{figure}

Consider an $N_t \times N_r$ massive MIMO channel with $O_t$ oscillators at the transmitter (corresponding to all users) and $O_r$ oscillators at the receiver. Each transmit-side oscillator feeds $N_{o,t} = N_t / O_t$ antennas, while similarly $N_{o,r} = N_r / O_r$ receive antennas are fed by the same oscillator\footnote{Notice that an even more general scenario in which different oscillators may feed different numbers of antennas is compatible with our proposal. However, we chose not to pursue this case to avoid an unnecessary notation burden.}, as shown in Fig. \ref{fig:System}, where the general model for the MIMO channel and oscillator setup is detailed. Different oscillators introduce independent PN processes. The input-output relationship for the MIMO setup at time $n=1,2,\,\dots$ is given by
\begin{equation} \label{eq:cha}
\mathbf{y}[n] = \mathbf{\Phi}_R[n]\, \mathbf{H}[n]\, \mathbf{\Phi}_T[n] \mathbf{x}[n] +
\mathbf{z}[n],
\end{equation}
where:
\begin{itemize}
\item $\mathbf{H}[n]$ is the $N_r \times N_t$ channel matrix at time $n$, whose statistical characterization will be described in more detail below and in Subsection \ref{sec:MSEresults};

\item $\mathbf{\Phi}_T[n] = \mathrm{diag}\left(e^{j\phi_1[n]}, \dots,  e^{j\phi_{O_t}[n]}\right) \otimes \Id_{N_{o,t}}$ and, similarly, $\mathbf{\Phi}_R[n] = \mathrm{diag}\left(e^{j\phi_{O_t+1}[n]}, \dots,  e^{j\phi_{O_t+O_r}[n]}\right)  \otimes \Id_{N_{o,r}}$ are the diagonal matrices of transmit and receive phase-noise coefficients at time $n$, respectively, assumed to be unknown at both sides; notice that we have assumed that  oscillator $i$, $i = 1,\dots,O_t$ feeds transmit antennas $(i-1)N_{o,t} + j$, $j = 1,\dots,N_{o,t}$, while oscillator $O_t + i$,  $i = 1,\dots,O_r$ feeds receive antennas $(i-1)N_{o,r} + j$, $j = 1,\dots,N_{o,r}$;

\item $\mathbf{x}[n]$ is the column vector of the $N_t$ modulated symbols transmitted at time $n$, each of them having average energy $E_s$;

\item $\mathbf{y}[n]$ is the column vector of the $N_r$ received samples at time $n$;

\item $\mathbf{z}[n]$ is a size-$N_r$ vector of zero-mean, circularly-invariant Gaussian-noise samples, with variance $\sigma^2$ per real dimension, which are assumed to be independent across time and for each receive antenna.
\end{itemize}

For the phase-noise samples, time dependency is kept into account by assuming Wiener phase-noise processes:
\begin{equation} \label{eq:wiener}
\phi_i[n] = \phi_i[n-1] + w_i[n], \,\,\,i=1,\dots,O_r+O_t,\,n=1,2,\dots
\end{equation}
where $\phi_1[0],\dots,\phi_{O_r+O_t}[0]$ are independent and uniformly distributed in $[0,2\pi)$, and $w_i[n],\dots,w_{O_r+O_t}[n]$ are independent zero-mean white Gaussian processes with power $\rho^2$ (all processes have the same power). 

Before describing the proposed detector, let us notice that each tap of the MIMO channel is affected by the sum of one transmit and one receive phase-noise process. We define a \emph{sum} phase-noise process as:
\begin{equation}
\phi_{ii'}[n]  = 
{\phi}_{i}[n] +
{\phi}_{O_t+i'}[n] ,\,\,\, i=1,\dots,O_t,\,i'=1,\dots,O_r .
\end{equation} 
Unlike the ${\phi}_{i}[n]$ processes, which will be called \emph{atomic} hereafter, two sum phase-noise processes are correlated if they share one index. Moreover, they can all be written as linear functions of $\phi_{i1}[n],\,\,i=1,\dots,O_t$ and $\phi_{1i'}[n],\,\,i'=2,\dots,O_r$, as:
 \begin{equation}
\phi_{ii'}[n]  = \phi_{i1}[n]+
\phi_{1i'}[n] -
\phi_{11}[n]. 
\end{equation}
Thus, the whole set of $O_r O_t$ sum phase-noise processes is generated from a basis with $O_r+O_t-1$ elements. Although atomic phase-noise processes have a simpler statistical characterization, they are not observable, so that phase estimation must pass through sum phase-noise process estimation. 

We define for future use the size-$(O_t+O_r)$ vector $\phiv[n]$, whose $i$-th element is $\phi_i[n]$, $i=1,\dots, O_r+O_t$. Analogously, we define the size-$(O_tO_r)$ vector $\phiv^{\mathrm{sum}}[n]$, whose element $(i-1)O_r + i'$ is $\phi_{ii'}[n]$, $i=1,\dots, O_t$, $i'=1,\dots, O_r$. 

\subsection{Pilot transmission scheme and channel estimation}

\begin{figure}[!ht]
\centering
\centerline{\includegraphics[width=\figurewidth]{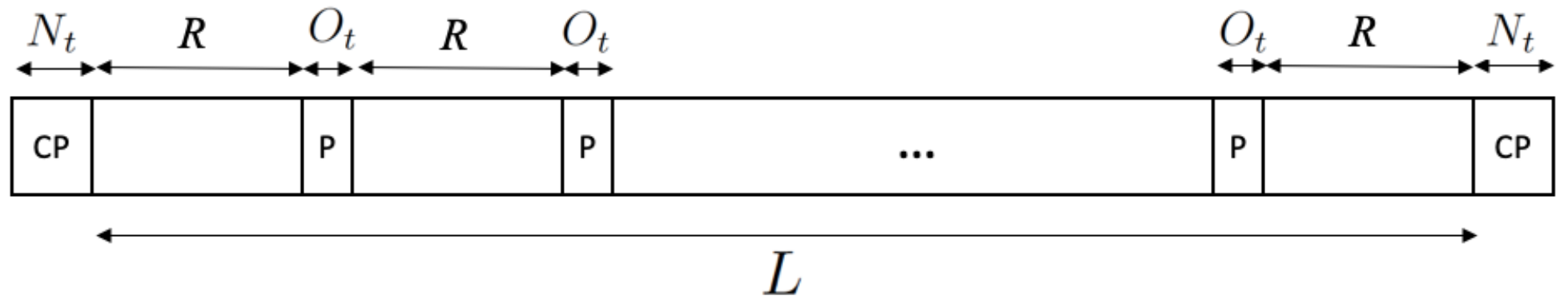}}
\caption{Frame structure for the proposed massive MIMO transmission scheme. CP: channel estimation pilots. P: phase-noise estimation pilots.}
\label{fig:Frame_structure}
\end{figure}

Fig. \ref{fig:Frame_structure} shows the structure of the frame assumed in this paper, with the diverse pilot and data transmission intervals, as described below.

The underlying hypothesis behind the channel model in \eqref{eq:cha} is that the channel matrix $\Hm[n]$ changes in time much more slowly than the PN. Thus, channel estimation can be performed less frequently than phase detection. In particular, we will suppose that there are two kinds of pilots in the system, with the first pilot type devoted to joint channel and phase estimation (as it is not possible, in principle, to distinguish the phase of unknown channel matrix entries from PN), and the second devoted only to phase detection.

The first type of pilots, called hereafter channel pilots, will be transmitted during $N_t$ consecutive channel uses, after which a frame of $L$ channel uses begins. We suppose \emph{orthogonal} channel pilots,  namely,  in the $i$-th channel use,  $i=1,\dots,N_t$,  all transmit antennas are switched off, except the $i$-th one. The average energy of  channel pilots will  be denoted $E_C$. The overall average energy of transmitted symbols will then be
\beq
\overline{E} = \frac{L}{L + N_t} E_s + \frac{N_t}{L + N_t} E_C.
\eeq
 
During such frame, we will make the assumption that the channel matrix stays constant (corresponding to a non-fading or block-fading scenario), so that we will simply write $\Hm$ instead of $\Hm[n]$. Moreover, different frames are treated independently, so that, for our purposes, we can consider a single frame, where channel estimation is performed at the beginning, say at step 0. As a result of channel estimation, the receiver will obtain a noisy version of $\Hm$, i.e.,\footnote{Here, we neglect PN variations during the channel estimation phase.}
\beq
\widehat{\Hm} = \mathbf{\Phi}_R[0]\, \mathbf{H}\, \mathbf{\Phi}_T[0] + \Zm_C,
\eeq
where $\Zm_C$ is the channel estimation error, assumed to be composed of i.i.d.
zero-mean Gaussian RV's with variance $\sigma^2/E_C$ per real dimension. In our model, we can always embed the initial phase-noise values into the channel matrix and assume that the phase-noise processes start from zero, or, equivalently, that we estimate differential processes $\phi_i[n] - \phi_i[0]$, $i = 1, \dots, O_r+O_t$.  With this in mind, we can redefine the channel matrix to be $\widetilde{\Hm} =  \mathbf{\Phi}_R[0]\, \mathbf{H}\, \mathbf{\Phi}_T[0]$.

The second type of pilots, used within the frame only to perform coarse phase detection, consists of blocks of $O_{t}$ pilots, transmitted every $R$ data symbol periods. In a given block, the $i$-th pilot, $i=1, \dots, O_{t}$,  is obtained by switching off all transmit antennas but the ones that are fed by the $i$-th transmit oscillator, which send symbols known at the receiver. Such phase pilots are used to obtain a rough estimate of sum phase-noise processes at pilot positions, after which MMSE filtering is performed to yield an estimate of \emph{atomic} phase-noise processes. Finally, linear interpolation allows to obtain an initial  phase estimate also for data positions, which will constitute a starting point for the EM-based receiver described in the next subsection. For more details on the EM initialization, see \cite{Tarable13}.

\section{The EM-based receiver}

We assume the information transmitted by the $k$-th user, $k=1,\dots,K$, is encoded with a channel encoder. The codeword is then mapped to a stream of QAM symbols, belonging to an $M$-size constellation $\left\{ \mathbf{x}^{(p)} \right\}_{p=1}^M$, and transmitted to the BS. Let us collect all symbols transmitted by all users into a frame of $L$ symbol vectors $\mathbf{X}=(\mathbf{x}[1],\dots,\mathbf{x}[L])$. Each symbol vector is transmitted through the channel described by equation (\ref{eq:cha}). The channel output for a frame is then collected in the matrix $\Ym = (\yv[1],\dots,\yv[L])$.

\begin{figure*}[!ht]
\centering
\centerline{\includegraphics[width=0.8\textwidth]{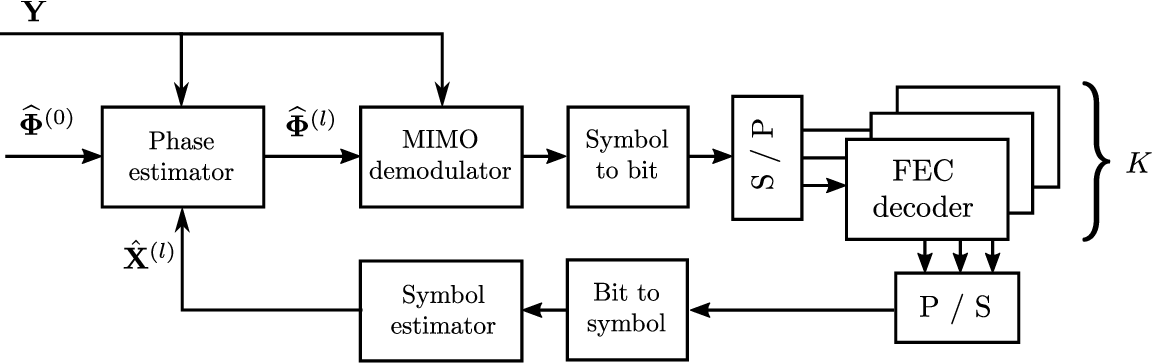}}
\caption{Structure of the iterative receiver.}
\label{fig:RX_scheme}
\end{figure*}

Let us define $f_A(\mathbf{\Phi})$ as the a priori distribution of $\mathbf{\Phi} = (\phiv[1],\dots,\phiv[L])$. The optimal joint phase detector and demodulator obtains the maximum a posteriori estimate of $\mathbf{\Phi}$ and $\Xm$, i.e.,  
\begin{eqnarray}
(\widehat{\Xm}, \widehat{\Phim}) & = & \arg \max_{\Xm, \Phim} \log \Pr \{\Xm,  \Phim | \Ym\} \\
& = & \arg \max_{\Xm, \Phim} \left( \log \Pr \{\Ym | \Xm,  \Phim\} + \log f_A\left(\mathbf{\Phi}\right) \right) 
\end{eqnarray}

As the solution of the above optimization problem is way too complex to be computed as is, the proposed receiver approximates it by an iterative alternate optimization, as follows. Let us start from an initial symbol distribution in which all transmitted symbols are independently and uniformly distributed on the constellation alphabet, i.e.,
\[
{\Pr}^{(0)}\{ \Xm\} = \frac1{M^{L N_t}}
\] 
Then, at iteration $l = 1,2,\dots$, perform the following computations:
\ifonecol
\begin{eqnarray}
\widehat{\mathbf{\Phi}}^{(l)} = (\widehat{\phiv}^{(l)}[1],\dots,\widehat{\phiv}^{(l)}[L]) &=& \arg \max_{\mathbf{\Phi}}  \left(\EE_{\Xm}^{(l-1)}\sum_{n=1}^{L}    \log\Pr\left\{ \mathbf{y}[n] |  \mathbf{x}[n] , \phiv[n]\right\} + \log f_A(\mathbf{\Phi})\right) \label{eq:Phihat} \\
\widehat{\mathbf{X}}^{(l)} = (\widehat{\xv}^{(l)}[1], \dots , \widehat{\xv}^{(l)}[L]) &=& \arg \max_{\mathbf{X}}  \left( \sum_{n=1}^{L} \log \Pr\left\{ \mathbf{y}[n] |  \mathbf{x}[n] , \widehat{\phiv}^{(l)}[n]\right\}\right),
\end{eqnarray}
\else
\begin{eqnarray}
\widehat{\mathbf{\Phi}}^{(l)}\!\!\!\!\! &=& \!\!\!\!\!\arg \max_{\mathbf{\Phi}}  \left(\EE_{\Xm}^{(l-1)}\log \Pr \{\Ym | \Xm,  \Phim\}+ \log f_A(\mathbf{\Phi})\right) 
\label{eq:Phihat}\\
\widehat{\mathbf{X}}^{(l)}\!\!\!\!\! &=&  \!\!\!\!\!\arg \max_{\mathbf{X}}  \left( \log \Pr \left\{\Ym | \Xm,  \widehat{\mathbf{\Phi}}^{(l)}\right\}\right),
\end{eqnarray}
\fi
where $\EE_{\Xm}^{(l)}$ represents the average with respect to the prior distribution ${\Pr}^{(0)}\{ \Xm\}$ for $l = 0$, and with respect to the posterior distribution ${\Pr}\{ \Xm | \Ym , \widehat{\mathbf{\Phi}}^{(l)} \}$ for $l > 0$.  The first step in the above alternate optimization is also called \emph{expectation-maximization} (EM) algorithm and is a typical suboptimal approach to MAP (or ML) estimation of unknown parameters such as,  in our case,  the phase noise. 

In practice, both steps in the above alternate maximization are approximated. The block diagram of the proposed receiver can be seen in Fig. \ref{fig:RX_scheme}. Starting from an initial pilot-based phase estimate $\widehat{\mathbf{\Phi}}^{\left(0\right)}$, the receiver obtains at the $l-$th iteration the phase estimate  $\widehat{\mathbf{\Phi}}^{\left(l\right)}$. The latter is then used in the MIMO demodulator to compute log-likelihood ratios (LLRs) for the transmitted symbols. These are converted to bit LLRs and input to the FEC decoders, one for each user. The decoders' output is then mapped into estimated transmitted symbols $\widehat{\mathbf{X}}^{\left(l\right)}$ and fed back to the phase estimator for the subsequent $\left(l+1\right)-$th iteration. In the first iteration, there is no prior information about the estimated received symbols. In Subsection \ref{sec:phase_estimator}, we will explain in detail the phase estimation algorithm, while in Subsection \ref{sec:mimo_dem} we briefly describe the MIMO demodulator.

\subsection{Phase estimator \label{sec:phase_estimator}} 

We open this subsection with some notation definition. We will denote as $\widetilde{\Hm}_{ij}$, $i = 1,\dots,O_r$, $j = 1,\dots,O_t$,  the $ N_{o,r} \times N_{o,t}$ block of $\widetilde{\Hm}$ corresponding to oscillators $j$ and $O_t + i$ at the transmitter and receiver side, respectively. Moreover, $\widetilde{\Hm}_{i}$, $i=1,\dots,O_t$ will be the $N_r \times N_{o,t}$ matrix obtained by stacking all matrices $\widetilde{\Hm}_{ji}$ for $j = 1,\dots, O_r$. Finally, we will denote as $\widetilde{\Hm}_{O_t + i}$, $i=1,\dots,O_r$ the $N_{o,r} \times N_t$ matrix obtained by juxtaposing all matrices $\widetilde{\Hm}_{ij}$ for $j = 1,\dots, O_t$. With $\widehat{\Hm}_{ij}$, $\widehat{\Hm}_{i}$ and $\widehat{\Hm}_{O_t + i}$ we will denote the corresponding estimated submatrices.

As already said, phase detection is based on an EM approach, where, at a given iteration, the expectation is taken with respect to the current distribution of the transmitted symbols. To ease notation, let us write
\begin{eqnarray} 
h^{(l)}(\mathbf{\Phi})&=&\EE_{\Xm}^{(l-1)}\log \Pr \{\Ym | \Xm,  \Phim\} \non
&=& \EE_{\mathbf{X}}^{(l-1)} \log \prod_{n=1}^{L}\Pr\left\{ \mathbf{y}[n] |  \mathbf{x}[n] , \phiv[n]\right\}.\label{eq:Estep}
\end{eqnarray}

For the channel model in (\ref{eq:cha}), apart from an inessential additive constant, (\ref{eq:Estep}) becomes:
\ifonecol
\begin{equation} \label{eq:Estep1}
h^{(l)}(\mathbf{\Phi})= \EE_{\mathbf{X}}^{(l-1)}  \sum_{n=1}^{L}  \frac{\Re\{\mathbf{x}^H[n] \mathbf{\Phi}_T^H[n] 
\widehat{\mathbf{H}}^H \mathbf{\Phi}_R^H[n] \mathbf{y}[n]\} - \frac1{2} \|\widehat{\mathbf{H}} \mathbf{\Phi}_T[n]\mathbf{x}[n]\|^2}{\sigma^2 \left( 1+E_C^{-1} \xv^H[n] \xv[n] \right)},
\end{equation}
\else
\begin{eqnarray} \label{eq:Estep1}
&\displaystyle h^{(l)}(\mathbf{\Phi}) \! = \EE_{\mathbf{X}}^{(l-1)} \!\! \sum_{n=1}^{L}\!\! \left(  \frac{\Re\{\mathbf{x}^H[n] \mathbf{\Phi}_T^H[n] 
\widehat{\mathbf{H}}^H \mathbf{\Phi}_R^H[n] \mathbf{y}[n]\}}{\sigma^2 \left( 1+E_C^{-1} \xv^H[n] \xv[n] \right)} -\right.& \nonumber \\
&\left.\displaystyle \frac{\frac1{2} \|\widehat{\mathbf{H}} \mathbf{\Phi}_T[n]\mathbf{x}[n]\|^2}{\sigma^2 \left( 1+E_C^{-1} \xv^H[n] \xv[n] \right)}\right),&
\end{eqnarray}
\fi
where the two terms in the denominator account for both the additive white Gaussian noise and the noisy channel estimation. It is shown in \cite{Tarable13} that there is no substantial performance loss if we substitute the average on $\Xm$ with the hard estimate of the transmitted symbols at iteration $l$. Thus, we make the following approximation:
\ifonecol
\begin{equation} \label{eq:Estep1_approx}
h^{(l)}(\mathbf{\Phi})\simeq  \sum_{n=1}^{L}  \frac{\Re\{(\widehat{\xv}^{(l)}[n])^H \mathbf{\Phi}_T^H[n] 
\widehat{\mathbf{H}}^H \mathbf{\Phi}_R^H[n] \mathbf{y}[n]\} - \frac1{2} \|\widehat{\mathbf{H}} \mathbf{\Phi}_T[n]\widehat{\xv}^{(l)}[n]\|^2}{\sigma^2 \left( 1+E_C^{-1} (\widehat{\xv}^{(l)}[n])^H \widehat{\xv}^{(l)}[n] \right)}.
\end{equation}
\else
\begin{eqnarray} \label{eq:Estep1_approx}
&\displaystyle h^{(l)}(\mathbf{\Phi}) \simeq \sum_{n=1}^{L} \left(  \frac{\Re\{(\widehat{\xv}^{(l)}[n])^H \mathbf{\Phi}_T^H[n] \widehat{\mathbf{H}}^H \mathbf{\Phi}_R^H[n] \mathbf{y}[n]\}}{\sigma^2 \left( 1+E_C^{-1} (\widehat{\xv}^{(l)}[n])^H \widehat{\xv}^{(l)}[n] \right)} - \right.& \nonumber \\ 
&\displaystyle \left.\frac{\frac1{2} \|\widehat{\mathbf{H}} \mathbf{\Phi}_T[n]\widehat{\xv}^{(l)}[n]\|^2}{\sigma^2 \left( 1+E_C^{-1} (\widehat{\xv}^{(l)}[n])^H \widehat{\xv}^{(l)}[n] \right)}\right).  &
\end{eqnarray}
\fi
where $\widehat{\xv}^{(l)}[n]$ is the $n$-th column of $\widehat{\mathbf{X}}^{(l)}$, i.e.,  the estimate of $\xv[n]$ at iteration $l$ (which is computed at iteration $l-1$ for $l>1$, while is taken to be zero for $l=1$).

In practice, as in \cite{DauKoLoe05a}, the maximization involved in \eqref{eq:Phihat} is approximated through the steepest-descent algorithm, and only the gradient of function $h^{(l)}(\mathbf{\Phi})$ is computed in the phase estimator. Let $\widehat{\mathbf{\Phi}}^{(l)}_m$ be the estimate of $\widehat{\mathbf{\Phi}}^{(l)}$ after $m$ steepest-descent iterations (starting from $\widehat{\mathbf{\Phi}}^{(l)}_0 = \widehat{\mathbf{\Phi}}^{(l-1)}$, i.e. the initial rough estimate based on phase pilots for $l=1$, the estimate at the previous iteration for $l>1$). Then:
\begin{equation}
\widehat{\mathbf{\Phi}}^{(l)}_{m} = \widehat{\mathbf{\Phi}}^{(l)}_{m-1} + \lambda^{(l)}_m \nabla_{\mathbf{\Phi}}
\left( h^{(l)}(\mathbf{\Phi}) + \log f_A(\mathbf{\Phi}) \right)\Bigg|_{\widehat{\mathbf{\Phi}}^{(l)}_{m-1}},
\end{equation}
where $\lambda^{(l)}_m$ is the $m$-th step size of the steepest-descent algorithm at iteration $l$.  The derivative of $h^{(l)}(\mathbf{\Phi})$ with respect to $\phi_i[n]$, $i=1,\dots,O_t$, is readily computed as:
\ifonecol
\begin{equation} \label{eq:deri1}
\frac{\partial}{\partial \phi_i[n]} h^{(l)}(\mathbf{\Phi}) \simeq \frac{\Im\{e^{-j \phi_i[n]}(\widehat{\xv}_i^{(l)}[n] )^H \widehat{\mathbf{H}}_i^H (\mathbf{\Phi}_R^H[n] \mathbf{y}[n] - \widehat{\mathbf{H}} \mathbf{\Phi}_T[n] \widehat{\mathbf{x}}^{(l)}[n])\}}{\sigma^2 \left( 1+E_C^{-1} (\widehat{\xv}^{(l)}[n])^H \widehat{\xv}^{(l)}[n] \right)},
\end{equation}
\else
\begin{eqnarray} \label{eq:deri1}
&\displaystyle \frac{\partial}{\partial \phi_i[n]} h^{(l)}(\mathbf{\Phi}) & \\
&\displaystyle \!\!\!\simeq\!\! \frac{\Im\{\!e^{-j \phi_i[n]}(\widehat{\xv}_i^{(l)}\![n] )^H \widehat{\mathbf{H}}_i^H \! (\mathbf{\Phi}_R^H[n] \mathbf{y}[n] \!-\! \widehat{\mathbf{H}} \mathbf{\Phi}_T[n] \widehat{\mathbf{x}}^{(l)}\![n])\!\}}{\sigma^2 \left( 1+E_C^{-1} (\widehat{\xv}^{(l)}[n])^H \widehat{\xv}^{(l)}[n] \right)}, & \nonumber
\end{eqnarray}
\fi
with $\widehat{\xv}_i^{(l)}[n]$ the length-$N_{o,t}$ subvector of $\widehat{\xv}^{(l)}[n]$ corresponding to transmit oscillator $i$. Analogously, the derivative of $h^{(l)}(\mathbf{\Phi})$ with respect to $\phi_{O_t+i}[n]$, $i=1,\dots,O_r$, results:
\ifonecol
\begin{equation} \label{eq:deri2}
\frac{\partial}{\partial \phi_{O_t+i}[n]} h^{(l)}(\mathbf{\Phi}) \simeq \frac{\Im\{e^{-j \phi_{O_t+i}[n]}(\widehat{\xv}^{(l)}[n])^H \mathbf{\Phi}_T^H[n] \widehat{\mathbf{H}}_{O_t+i}^H  \mathbf{y}_i[n]\}}{\sigma^2 \left( 1+E_C^{-1} (\widehat{\xv}^{(l)}[n])^H \widehat{\xv}^{(l)}[n] \right)},
\end{equation}
\else
\begin{eqnarray} \label{eq:deri2}
&\displaystyle \frac{\partial}{\partial \phi_{O_t+i}[n]} h^{(l)}(\mathbf{\Phi}) & \\
&\displaystyle \simeq \frac{\Im\{e^{-j \phi_{O_t+i}[n]}(\widehat{\xv}^{(l)}[n])^H \mathbf{\Phi}_T^H[n] \widehat{\mathbf{H}}_{O_t+i}^H  \mathbf{y}_i[n]\}}{\sigma^2 \left( 1+E_C^{-1} (\widehat{\xv}^{(l)}[n])^H \widehat{\xv}^{(l)}[n] \right)},& \nonumber
\end{eqnarray}
\fi
with $\mathbf{y}_i[n]$ the length-$N_{o,r}$ subvector of $\mathbf{y}[n]$ corresponding to receiver oscillator $O_t+i$. Moreover, thanks to the Wiener model, we have (with a slight abuse of notation):
\begin{equation}
\log f_A(\mathbf{\Phi}) = \log f_A(\mathbf{\phi}[0]) + \sum_{n=1}^{L} \log f_A(\mathbf{\phi}[n+1] | \mathbf{\phi}[n]),
\end{equation}
so that, provided that $\rho^2 \ll 2\pi$ (a usually realistic approximation), we obtain \cite{DauKoLoe05a}:
\begin{eqnarray}
\frac{\partial}{\partial \phi_i[n]} \log f_A(\mathbf{\Phi}) &=& \frac{\partial}{\partial \phi_i[n]} \log f_A(\mathbf{\phi}[n] | \mathbf{\phi}[n-1])+ \nonumber\\
&& \frac{\partial}{\partial \phi_i[n]} \log f_A(\mathbf{\phi}[n+1] | \mathbf{\phi}[n]) \nonumber \\
& \simeq & \frac{\phi_i[n-1]-\phi_i[n]+k_{n-1}2\pi}{\rho^2}+ \nonumber\\
&&  \frac{\phi_i[n+1]-\phi_i[n]+k_{n}2\pi}{\rho^2} \label{eq:ap},
\end{eqnarray}
where $k_n$ and $k_{n-1}$ are signed integers that minimize the moduli of the numerators.

\subsubsection{Step size optimization and stopping rule}

The performance of the steepest-descent algorithm depends crucially on the choice of the step size and on the number of performed steps. In this subsection, we clarify the design solutions we propose in this regard.

Regarding the step size, when it is fixed along the steps, it may imply a very slow convergence, if it is chosen too low, or even a lack of convergence to the optimum point, if it chosen too large. In order to improve the convergence of the steepest-descent algorithm, we have then dynamically chosen the step size $\lambda^{(l)}_m$ as follows.
\begin{itemize}
\item In the first step ($m=1$), we adopt the backtracking line search approach, which has some convergence guarantees and it is popular also in deep-learning literature \cite{Truong21}. We start from a maximum candidate value for $\lambda^{(l)}_1$, and we progressively reduce it until Armijo's condition is satisfied \cite{Armijo66}, i.e.
\ifonecol
\beq
g\left(\widehat{\mathbf{\Phi}}^{(l)}_0 + \lambda^{(l)}_1 \nabla g\left(\widehat{\mathbf{\Phi}}^{(l)}_0\right) \right) \geq g\left(\widehat{\mathbf{\Phi}}^{(l)}_0\right) + \lambda^{(l)}_1 c \Big\|\nabla g\left(\widehat{\mathbf{\Phi}}^{(l)}_0\right) \Big\|^2,
\eeq
\else
\begin{equation}
 g\!\left(\widehat{\mathbf{\Phi}}^{(l)}_0 \!+\! \lambda^{(l)}_1 \nabla g\!\left(\widehat{\mathbf{\Phi}}^{(l)}_0\right)\! \right) \!\geq\! g\!\left(\widehat{\mathbf{\Phi}}^{(l)}_0\right) + \lambda^{(l)}_1 c \Big\|\nabla g\!\left(\widehat{\mathbf{\Phi}}^{(l)}_0\right)\!\! \Big\|^2,
\end{equation}
\fi
where $g(\mathbf{\Phi}) =  h^{(l)}(\mathbf{\Phi}) + \log f_A(\mathbf{\Phi}) $ and $c$ is a constant, set to $c = 0.5$ in our simulations.  The progressive reduction of $\lambda^{(l)}_1$ from the initial value is performed with steps of 0.5. 

\item In the subsequent iterations ($m>1$), to set the step size we use the  Barzilai-Borwein method \cite{Barzilai88}, which is simpler than backtracking line search and has also global-convergence guarantees under mild conditions, yielding:
\ifonecol
\beq
\lambda^{(l)}_m = \frac{\Big|\left(\widehat{\mathbf{\Phi}}^{(l)}_{m-1} - \widehat{\mathbf{\Phi}}^{(l)}_{m-2}\right)^T \cdot \left(\nabla g\left(\widehat{\mathbf{\Phi}}^{(l)}_{m-1}\right) - \nabla g\left(\widehat{\mathbf{\Phi}}^{(l)}_{m-2}\right)\right)\Big|}{\Big\|\nabla g\left(\widehat{\mathbf{\Phi}}^{(l)}_{m-1}\right) - \nabla g\left(\widehat{\mathbf{\Phi}}^{(l)}_{m-2}\right)\Big\|^2}.
\label{eq:BB}
\eeq
\else
\beq
\lambda^{(l)}_m \!=\! \frac{\Big|\!\!\left(\widehat{\mathbf{\Phi}}^{(l)}_{m-1}\!\! - \widehat{\mathbf{\Phi}}^{(l)}_{m-2}\right)^T\!\!\!\!\cdot\! \left(\nabla g\!\left(\widehat{\mathbf{\Phi}}^{(l)}_{m-1}\!\right) \! - \! \nabla g\!\left(\widehat{\mathbf{\Phi}}^{(l)}_{m-2}\!\right)\!\right)\!\!\Big|}{\Big\|\nabla g\left(\widehat{\mathbf{\Phi}}^{(l)}_{m-1}\right) - \nabla g\left(\widehat{\mathbf{\Phi}}^{(l)}_{m-2}\right)\Big\|^2}.
\label{eq:BB}
\eeq
\fi
\end{itemize}

The steepest-descent iterations are stopped whenever the relative increment in the objective function falls below a given threshold, i.e., whenever
\beq
\label{eq:grad_th}
\frac{\Big|g\left(\widehat{\mathbf{\Phi}}^{(l)}_{m}\right) - g\left(\widehat{\mathbf{\Phi}}^{(l)}_{m-1}\right)\Big|}{g\left(\widehat{\mathbf{\Phi}}^{(l)}_{m-1}\right)} < \theta.
\eeq
Setting an appropriate value for $\theta$ implies meeting a trade-off between convergence time and final error performance. A thorough study of the impact of the threshold on performance and complexity has been carried out and the results are presented in Sections \ref{sec:MSEresults} and \ref{sec:iterations}, respectively.

\subsection{MIMO demodulator \label{sec:mimo_dem}}

At a given receiver iteration, the block labelled ``MIMO demodulator'' in Fig. \ref{fig:RX_scheme} derives LLRs on symbols, based on the current phase-noise estimates, followed by a standard extraction of LLRs on coded bits. These bit LLRs are then input to the channel decoder.  Due to the large size of the massive MIMO system, we propose a suboptimal MIMO demodulator, based on linear minimum mean-square error (LMMSE) filtering. Explicitly, at iteration $l$ and channel use $n$, the symbol LLRs are computed on a per-symbol basis from the LMMSE filter output
\beq
\widetilde{\yv}[n] = \Fm^{(l)}[n] \yv[n],
\eeq
where
\beq
\Fm^{(l)}[n] = \left(\widehat{\Hm}^H \widehat{\Hm} + \sigma^2 \left( \frac{N_t}{E_C	} + \frac1{E_s} \right) \right)^{-1} \widehat{\Hm}^H.
\eeq
Notice that the above filter expression takes into account the effect of imperfect CSI.

\section{Bayesian Cramer-Rao bound}
\label{sec:Cramer}

In this section, we compute the BCRB for the system described in the previous section, under the hypothesis that the transmitted symbols are \emph{known} at the receiver and assuming perfect channel state information. This hypothesis suits well the case of the EM receiver, since, if convergence eventually occurs, the output of the channel decoder after several iterations provides the phase detector with (almost) perfect knowledge of the transmitted symbols. The subject of this subsection is the generalization of the BCRB computation in \cite{Nasir13, Tarable14}, which considers the particular case of $N_{o,t} = N_{o,r} = 1$.

The BCRB allows to lower-bound the MSE between the unknown phase-noise sample sequence and its estimate performed by the phase detector at the receiver. Let $\bar{\phiv}^{\mathrm{sum}}$ be the size-$O_t O_r L$ column vector obtained by stacking all vectors $\phiv^{\mathrm{sum}}[1], \dots, \phiv^{\mathrm{sum}}[L]$ and let $\widehat{\bar{\phiv}}^{\mathrm{sum}}$ be any possible estimate of $\bar{\phiv}^{\mathrm{sum}}$. The covariance matrix $\Sigmam$ of such estimate is given by
\beq
\Sigmam = \EE_{\Phim, \Ym, \Xm, \widehat{\Hm}} \left\{ \left(\widehat{\bar{\phiv}}^{\mathrm{sum}} - \bar{\phiv}^{\mathrm{sum}} \right) \cdot \left(\widehat{\bar{\phiv}}^{\mathrm{sum}} - \bar{\phiv}^{\mathrm{sum}} \right)\Tran \right\}.
\eeq 
The BCRB then states that
\beq
\Sigmam \succeq \Mm^{-1},
\eeq 
where $\Mm$ is the Bayesian information matrix (BIM) defined by 
\beq \label{eq:info_matrix}
\Mm = -\EE_{\Phim, \Ym, \Xm}  \left\{ \nabla_{\bar{\phiv}^{\mathrm{sum}}} \nabla_{\bar{\phiv}^{\mathrm{sum}}}\Tran \log f\left( \Ym, \bar{\phiv}^{\mathrm{sum}}|  \Xm\right) \right\}.
\eeq
Conditioning on $\Xm$ in the above definition corresponds to the hypothesis of known transmitted symbols. Considering the $i-$th element of vector $\bar{\phiv}^{\mathrm{sum}}$, the BCRB implies that:
\beq
\left( \Sigmam \right)_{i,i} \geq (\Mm^{-1})_{i,i},
\eeq
i.e., the BCRB provides lower bounds to the MSE for every estimator of the sum phase-noise samples.

The result is summarized in the following proposition. 

\begin{proposition}
For the channel model in \eqref{eq:cha}, if $\rho \ll 2\pi$, the BIM $\Mm$ defined in \eqref{eq:info_matrix} is given by
\beq \label{eq:BIM}
\Mm = \left[
\begin{array}{ccccc}
\Mm_0 & \Mm_1 & & &  \\
\Mm_1 & \Mm_0 & \Mm_1 & &  \\
& \Mm_1 & \Mm_0 & \ddots & \\
 & & \ddots & \ddots & \Mm_1 \\
 & & & \Mm_1 & \Mm_0
\end{array}
\right],
\eeq
where $\Mm_0 = \Jm\Tran \widetilde{\Mm}_0 \Jm$ and $\Mm_1 = \Jm\Tran \widetilde{\Mm}_1 \Jm$, $\Jm$ being the Jacobian of the transformation from sum processes to atomic processes at a given time $n$, $ \widetilde{\Mm}_1 = -\frac1{\rho^2} \Id_{O_t+O_r}$ and $ \widetilde{\Mm}_0 = \widetilde{\Mm}_0^Y + \frac{2}{\rho^2} \Id_{O_t+O_r}$, and
\beq \label{eq:M0Y}
 \widetilde{\Mm}_0^Y = \frac1{\sigma^2}\left[ 
 \begin{array}{cc}
 \Gammam_T & \Omegam\Tran \\
 \Omegam & \Gammam_R 
 \end{array}
 \right],
\eeq
where 
\[
\Gammam_T = \mathrm{diag}(\| \Hm_1 \|_F^2, \dots, \| \Hm_{O_t} \|_F^2) 
\] 
\[
\Gammam_R  = \mathrm{diag}(\| \Hm_{O_t+1} \|_F^2, \dots, \| \Hm_{O_t+O_r} \|_F^2) ,
\] 
 and $\Omegam$ is an $O_r \times O_t $ matrix whose $(i,j)$ entry is given by
 \[
\Omega_{ij} =  \| \Hm_{ij} \|_F^2 .
 \]
\end{proposition}

\pf
The formula is a rather straightforward generalization of those in \cite{Nasir13, Tarable14}.
\qedsymb

\section{Simulation results}
\label{Simulations}

In this section, we present comparative simulation results for some meaningful test cases. These have been performed for a scenario under the following specific conditions:
\begin{itemize}
 \item The modulation chosen is 64-QAM, with efficiency $6$ bits per symbol.
 \item The FEC scheme chosen is the 5G NR LDPC code of type $2$ and lift factor $Z = 2$, with a length $N = 104$ bits and a rate equal to $0.8$ \cite{5G}. The number of LDPC decoding iterations is set to $50$.
 \item The PN standard deviation is $\rho=0.2$ radians and the PN process model simulated is the Wiener one, unless explicitly stated.
 \item In general, we consider a Rician-fading channel matrix given by
\beq
\Hm = \sqrt{\frac{K_{Rice}}{K_{Rice}+1}} \Hm_{\mathrm{LOS}} + \sqrt{\frac1{K_{Rice}+1}} \Hm_{w}
\eeq
where $\Hm_{w}$ is composed by i.i.d. zero-mean circular complex Gaussian RV's with unit variance, and $\Hm_{\mathrm{LOS}}$ is composed by unit-modulus entries with i.i.d. phases uniformly distributed in $\left[0,2\pi\right)$. Unless otherwise stated, we set $K_{Rice} = 100$ dB, so that $\Hm =  \Hm_{\mathrm{LOS}}$.
 \item In the case of the simulations run to get BER statistics, a genie-aided decoding stopping rule is activated, in order to speed up the process. No such rule is activated in the case of the simulations run to get MSE statistics.
 \item The frame length takes value $L=1086$ symbols.
 \item The channel coherence time is $L$ symbol periods (block fading).
 \item At the transmitter side  $K=O_t=16$. Moreover, since we are considering an uplink massive MIMO for mobile communications, $N_{o,t}$ will be set to $2$ (two antennas for each user).
 \item At the receiver side, we set $N_r=64$, with $O_r=4$ (four oscillators at the RX side, feeding $N_{o,r}=16$ antennas each).
 \item Unless otherwise stated, the maximum number of receiver iterations is set to $10$.
 \item Each frame is built to contain $1000$ LDPC codewords.
 \item The simulations to obtain the BER results have been obtained for up to $100$ frames in error, with a maximum number of $10^6$ simulated frames.
 \item The maximum number of iterations in the steepest-descent algorithm has been set to $300$.
\end{itemize}

\subsection{MSE simulation results}
\label{sec:MSEresults}

\begin{figure}[!ht]
\centering
\centerline{\includegraphics[width=\figurewidth]{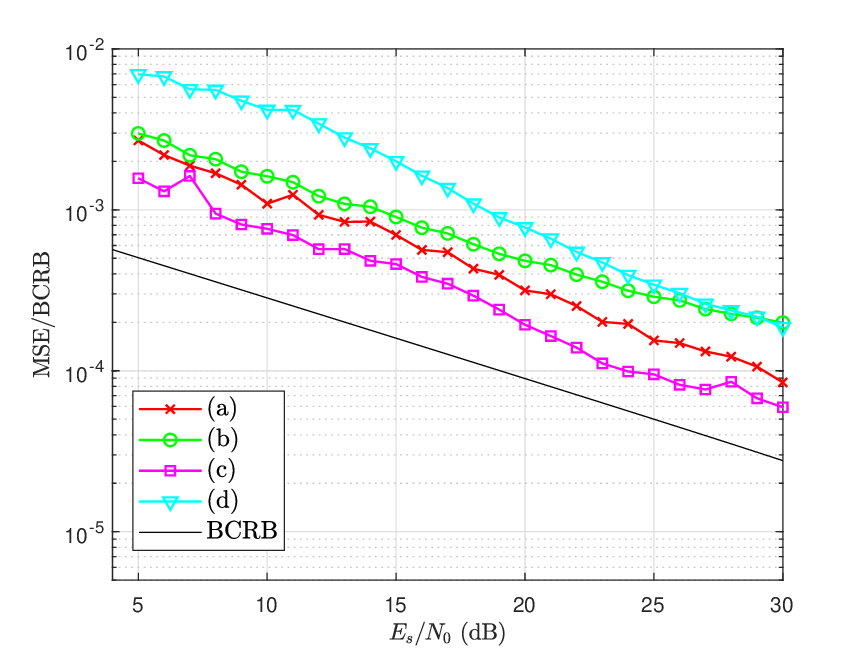}}
\caption{MSE results for perfect CSI. (a) Threshold $\theta=10^{-6}$ and $10$ receiver iterations. (b) Threshold $\theta=10^{-6}$ and $1$ receiver iteration. (c) Threshold $\theta=10^{-7}$ and $10$ receiver iterations. (d) Threshold $\theta=10^{-4}$ and $10$ receiver iterations.}
\label{fig:Fig1MSE}
\end{figure}

\begin{figure}[!ht]
\centering
\centerline{\includegraphics[width=\figurewidth]{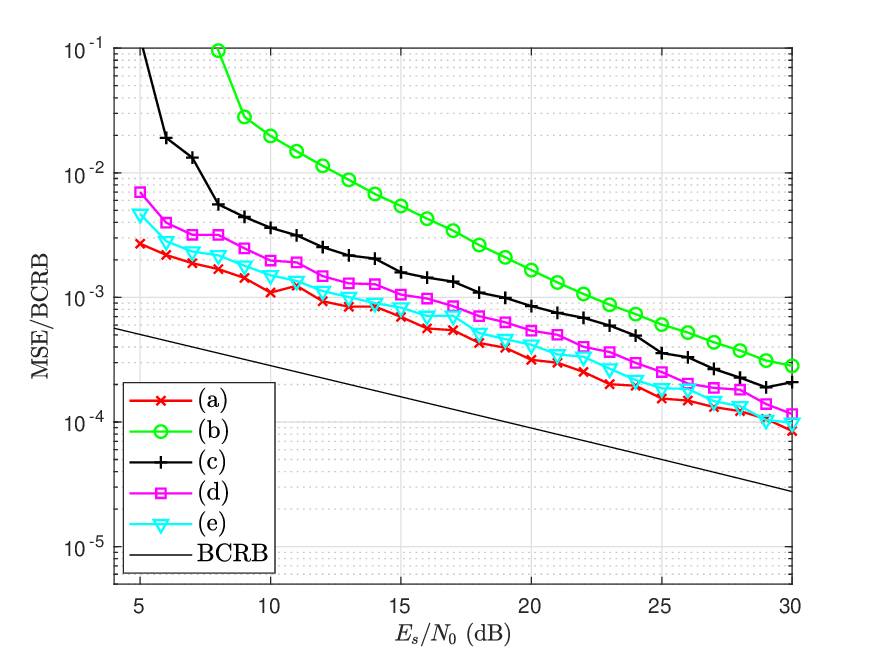}}
\caption{MSE results for perfect and non-perfect CSI. In all the cases, $\theta=10^{-6}$ and we have performed $10$ receiver iterations. (a) Perfect CSI. (b) $E_C/E_s=5$ dB. (c) $E_C/E_s=10$ dB. (d) $E_C/E_s=15$ dB. (e) $E_C/E_s=20$ dB.}
\label{fig:Fig2MSE}
\end{figure}

\begin{figure}[!ht]
\centering
\centerline{\includegraphics[width=\figurewidth]{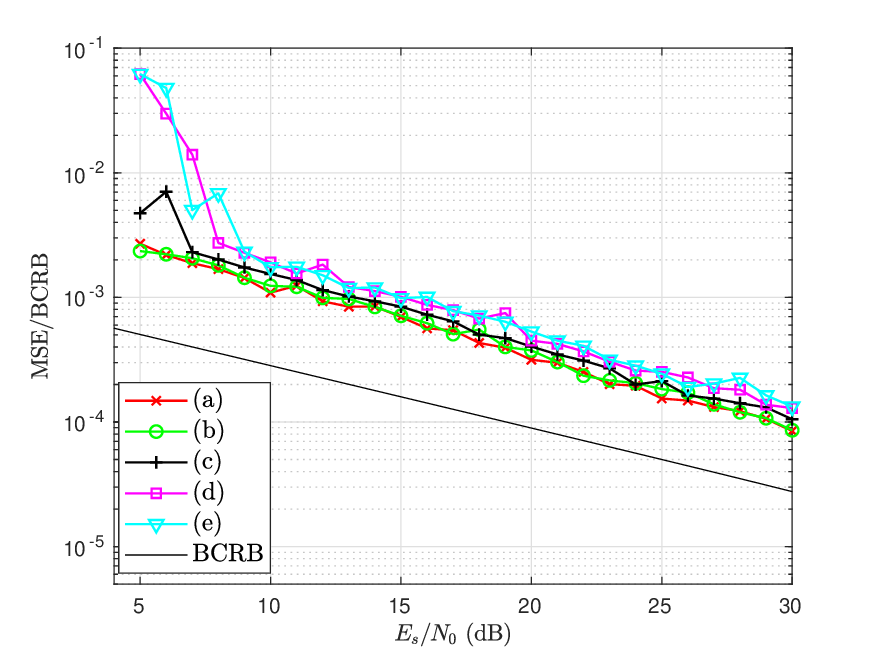}}
\caption{MSE results for the fading channel. BCRB is the same for AWGN and for every $K_{rice}$ factor. In all the cases, $\theta=10^{-6}$ and we have performed $10$ receiver iterations (a) $K_{Rice}=100$ dB. (b) $K_{Rice}=10$ dB. (c) $K_{Rice}=0$ dB. (d)  $K_{Rice}=-10$ dB. (e) $K_{Rice}=-100$ dB.}
\label{fig:Fig3MSE}
\end{figure}

Figs. \ref{fig:Fig1MSE}-\ref{fig:Fig3MSE} show the MSE of the phase estimation and the corresponding BCRB for several setups. In Fig. \ref{fig:Fig1MSE}, the curves show the MSE for different number of receiver iterations and several choices of the threshold $\theta$ to stop steepest-descent steps, as per \eqref{eq:grad_th}. The red curve shows the MSE for $\theta = 10^{-6}$ and $10$ receiver iterations. The benefit of the iterative receiver can be observed by comparing such curve with the green one, which corresponds to the MSE performance after one receiver iteration. Since the green curve has a lower slope, the benefit increases with the SNR, reaching about 7 dB for $\mathrm{MSE} = 2 \times 10^{-4}$. The magenta curve shows the MSE performance for $\theta = 10^{-7}$ and $10$ receiver iterations. The gain with respect to the red curve ranges from $2$ to $5$ dB but the slope of the two curves is about the same. However, as it will be evident from the results in Subsection \ref{BERresults}, when $\theta = 10^{-7}$, the average total number of steepest-descent steps is larger than for $\theta = 10^{-6}$, being for example $565$ for the former case and $263$ for the latter at $E_s/N_0 = 9$ dB. At an SNR large enough, the curves for $10$ iterations are essentially parallel to the BCRB, with an SNR gap at $\mathrm{MSE} = 10^{-4}$ of about $10$ dB for $\theta = 10^{-6}$ and $5$ dB for $\theta = 10^{-7}$.

In the following, for all simulation results, we have set a threshold  $\theta = 10^{-6}$ and $10$ receiver iterations. In Fig. \ref{fig:Fig2MSE}, we show the effect of imperfect channel CSI. The red curve shows the performance for perfect CSI (same as in Fig. \ref{fig:Fig1MSE}). The other curves show the performance with imperfect CSI and different channel pilot power. In particular, taking into account that $E_C$ is the energy of channel pilots, the curves from top to bottom correspond to a ratio $E_C/E_s$ equal to $5$, $10$, $15$ and $20$ dB, respectively. As it can be seen, the green curve is heavily affected at low-to-medium SNR values by the channel uncertainty. The black curve, which is obtained by supposing a power for pilots 10 times the power for payload, loses 5-6 dB with respect to perfect CSI, for $\mathrm{MSE} < 3 \times 10^{-4}$.

Fig. \ref{fig:Fig3MSE}  shows the effect of fading. The different curves are obtained with different values of the Rice factor, ranging from $K_{Rice} = 100$ dB, i.e., pure LOS, to $K_{Rice} = -100$ dB, i.e., Rayleigh fading. As it can be seen, above an SNR of $10$ dB, the MSE curves are only marginally affected by the fading intensity, with the MSE slightly increasing with decreasing $K_{Rice}$. It is worth noting that the BCRB is virtually independent of $K_{Rice}$, since, due to the large MIMO size and to the expression of the BIM in \eqref{eq:BIM}, an averaging effect takes place.

\subsection{BER simulation results}
\label{BERresults}

\begin{figure}[!ht]
\centering
\centerline{\includegraphics[width=\figurewidth]{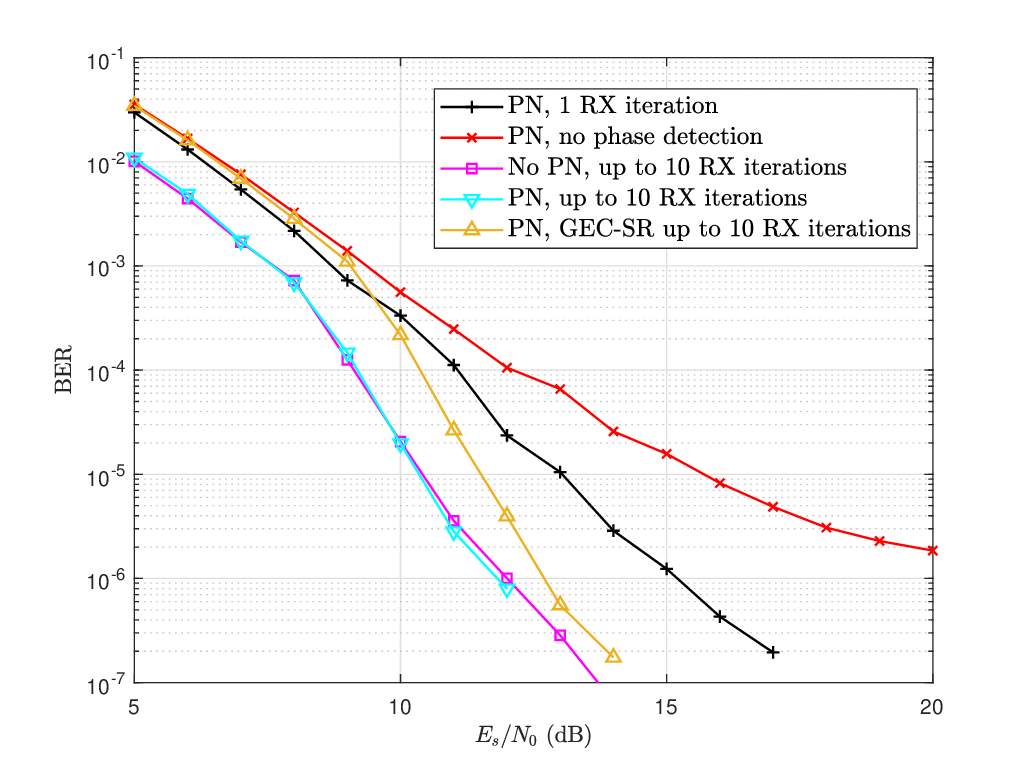}}
\caption{BER results for different cases with or without added PN (cases PN {\it vs.} No PN), and several receiver strategies (including no phase detection and a variable maximum number of iterations). The PN process has a standard deviation $\rho=0.2$ in all the PN-affected cases.}
\label{fig:Fig1BER}
\end{figure}

\begin{figure}[!ht]
\centering
\centerline{\includegraphics[width=\figurewidth]{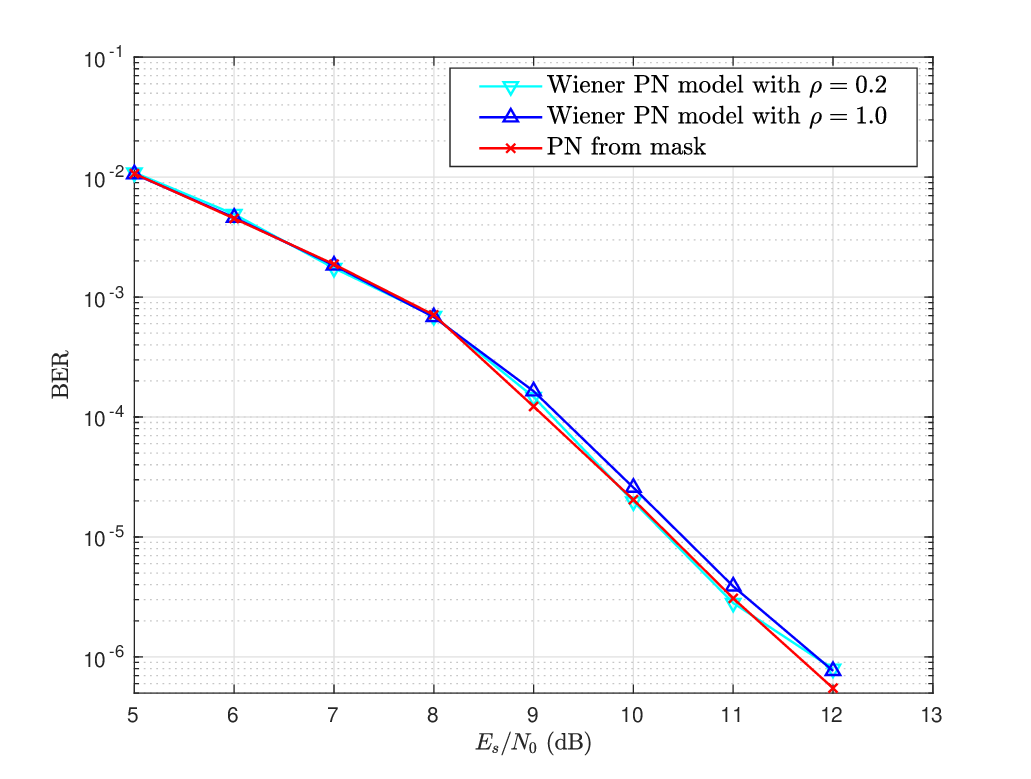}}
\caption{BER results with two different levels of PN power, along with the results obtained using PN samples produced from a mask instead of the Wiener model. The standard deviation of the mask-based PN process is approximately $0.2$.}
\label{fig:Fig2BER}
\end{figure}

\begin{figure}[!ht]
\centering
\centerline{\includegraphics[width=\figurewidth]{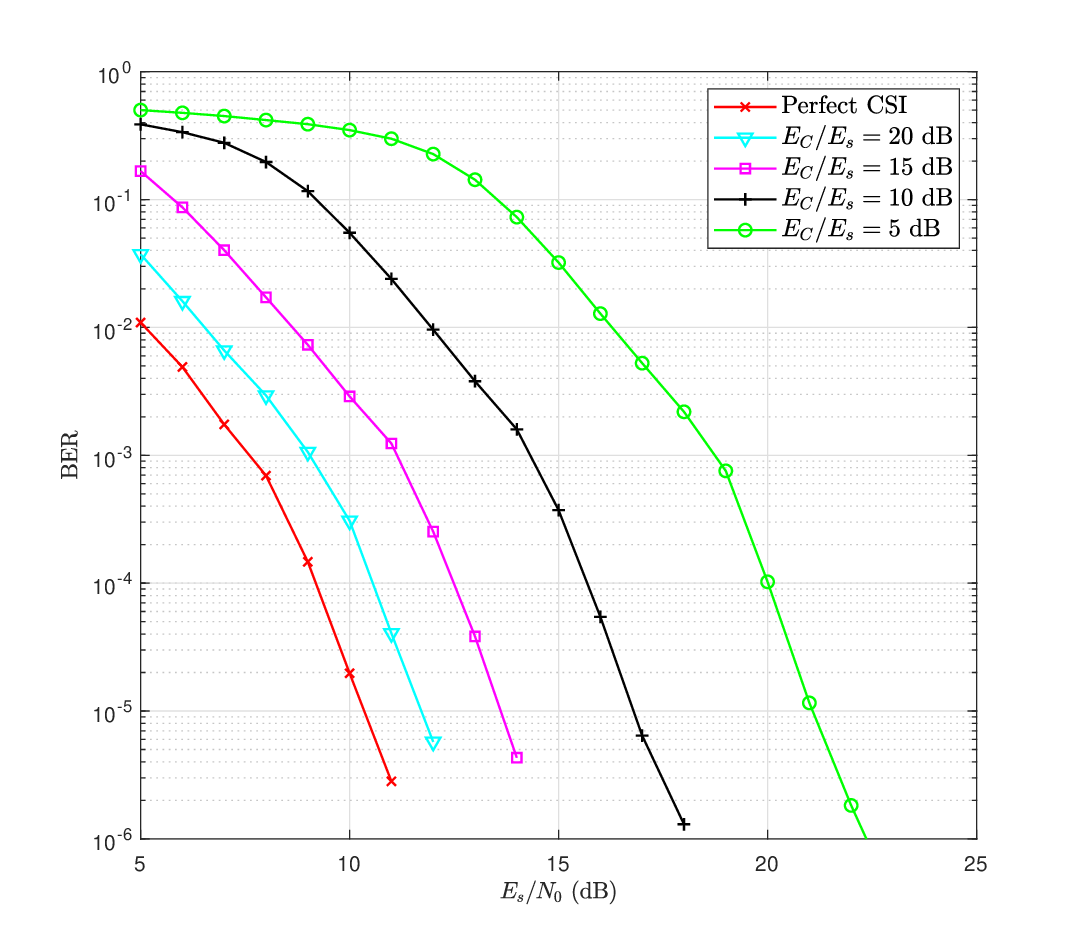}}
\caption{BER results for different cases with perfect and non-perfect CSI. In all the cases, the maximum number of receiver iterations is $10$, and the threshold for the steepest-descent algorithm is $\theta=10^{-6}$.} %
\label{fig:Fig3BER}
\end{figure}

\begin{figure}[!ht]
\centering
\centerline{\includegraphics[width=\figurewidth]{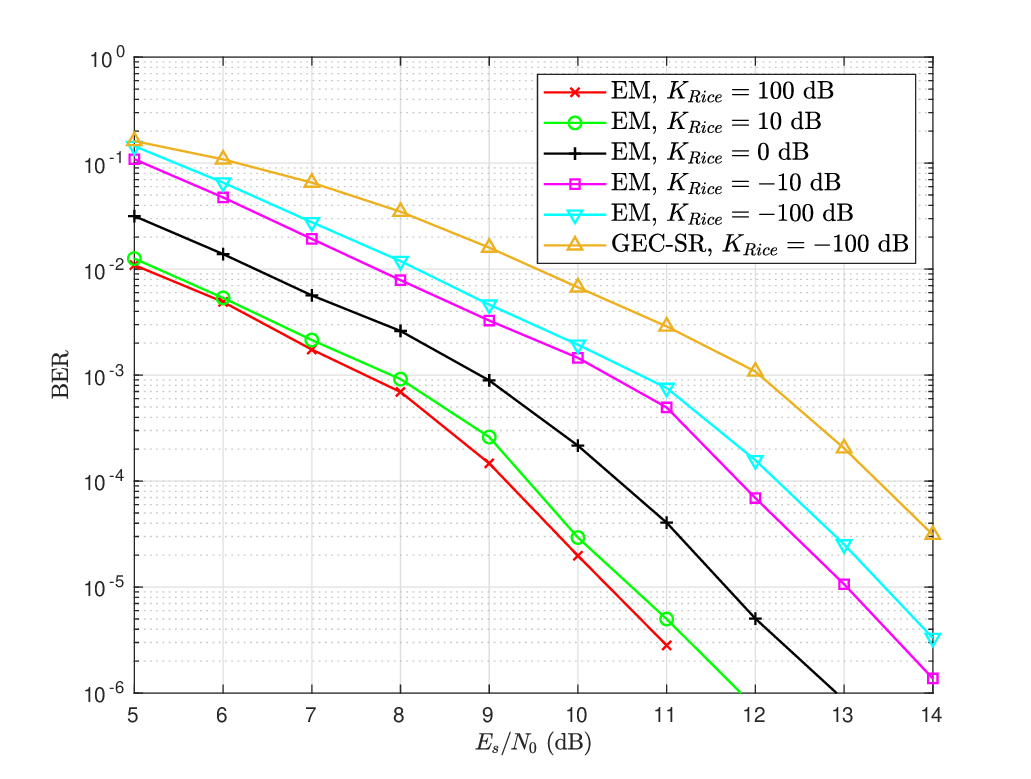}}
\caption{BER results for the fading channel, with $K_{Rice}$ ranging from $100$ to $-100$ dB. In all the cases with the EM algorithm, the maximum number of receiver iterations is $10$, and the threshold for the steepest-descent algorithm is $\theta=10^{-6}$. The results obtained for $K_{Rice}=-100$ dB with the above mentioned GEC-SR alternative are also shown for comparison.}
\label{fig:Fig4BER}
\end{figure}

In Figs. \ref{fig:Fig1BER} to \ref{fig:Fig4BER}, we have depicted the BER results for various parameters, keeping the number of antennas and oscillators of the basic setup used to illustrate the MSE behavior. In order to compare the BER performance of our phase detector with other alternative from the literature, we have also simulated a phase detector based on the so-called generalized expectation-consistent signal recovery (GEC-SR), as proposed in reference \cite{Yang19}. It is to be noted that, whereas the original detector developed in \cite{Yang19} is intended for symbol detection at each channel use for the case of one antenna per oscillator, we have adapted the algorithm therein to take into account the variable number of antennas per oscillator and to estimate the actual phase increments of the PN processes along the entire frame, in order to perform phase compensation prior to MIMO demodulation. Therefore, we are actually replacing the ``Phase Estimator'' block in Fig. \ref{fig:RX_scheme} with the GEC-SR-based alternative, while the rest of the iterative detector remains exactly the same. This ensures a fairer and more coherent comparison of the two phase-detection options.

In Fig. \ref{fig:Fig1BER}, we show the impact of PN on the system BER. First of all, the red curve depicts the BER for the case in which the PN has standard deviation $\rho = 0.2$ and there is no phase detection. As it can be seen, the curve gets flatter and flatter as the SNR increases, since for high SNR PN becomes the major and constant source of disturbance. If we employ a non-iterative receiver that performs phase estimation as described in the previous sections, the BER performance gets much better, as the black curve shows. However, the major improvement comes in with iterations, as shown by the light-blue curve, which corresponds to a maximum number of $10$ iterations. Once again, the proposed receiver is able to virtually achieve the BER of the case without PN, with the same number of iterations. The receiver without PN takes advantage of the iterations in the MIMO demodulator and in the symbol-to-bit demapper. We can compare these results with the ones obtained with the GEC-SR phase detector (orange curve). We can see how the GEC-SR algorithm exhibits a worse performance, only approaching the results of our proposal for high SNR. This is explained by the fact that the GEC-SR algorithm, as stated in \cite{Yang19}, relies on approximations that get more accurate as SNR grows. In any case, these results confirm the goodness of the EM-based phase estimator.

In Fig. \ref{fig:Fig2BER}, we can see the results with up to $10$ receiver iterations, and several possibilities respecting the PN process. In light blue, we have the BER curve from Fig. \ref{fig:Fig1BER}, when $\rho=0.2$. In dark blue, we have the results with the same setup, excepting a higher level in PN standard deviation, namely $\rho=1.0$. As it can be seen, with this level of PN power, the system is still able to approach the PN-free performance, with a slight worsening respecting $\rho=0.2$. In red, we have depicted the results obtained with the same setup, but in the case the PN process is generated from a mask, with an equivalent standard deviation of approximately $0.2$. The mask is characterized by slopes $-3$ dBc/decade, $-2$ dBc/decade and $0$ dBc/decade in the respective intervals $\left[2,100\right)$ KHz, $\left[0.1,1\right)$ MHz and $\left[1, 2 f_s\right)$ MHz, where the sampling frequency is $f_s=26$ MHz, and the reference level at $100$ KHz is $-133$ dBc/Hz. The mask captures better the PN process from a real-world practical point of view, and we can see that, for the same level of PN power, the results are compatible with the ones generated with the theoretical Winener PN model. The slight improvement could be mainly due to the fact that the actual standard deviation is a bit lower than $0.2$, along with the fact that the Wiener model represents a worst-case PN process model. These results show the robustness of the phase detector and receiver proposed, as it can cope with different levels of PN  power and with a more realistic PN process model.

From now on, we set $\rho=0.2$ and up to $10$ receiver iterations in the subsequent tests. In Fig. \ref{fig:Fig3BER}, we show the effect on BER of imperfect CSI. Like in Fig. \ref{fig:Fig2MSE}, we consider different relative power levels for channel pilots. For $\mathrm{BER} = 10^{-4}$, $E_C/E_s = 5$ dB (green curve) entails a $11$-dB loss with respect to the perfect-CSI case, while $E_C/E_s = 10$ dB (black curve) reduces the loss to about $8$ dB. When $E_C/E_s = 20$ dB (light-blue curve), the system is able to yield a limited loss of about $1.5$ dB. These gaps are in good accordance with those seen for the MSE, as can be stated from Fig. \ref{fig:Fig2MSE}.

In Fig. \ref{fig:Fig4BER}, we depict the results with perfect CSI but different $K_{Rice}$ values, displaying the same cases as in Fig. \ref{fig:Fig3MSE} for our proposed EM-based algorithm. As it can be verified, for a given performance, we have SNR differences very close to those shown in Fig. \ref{fig:Fig3MSE} respecting the MSE. For example, for a BER of $10^{-4}$, we can see the SNR gap between $K_{Rice}=100$ dB and $K_{Rice}=10$ dB is less than half a dB. The same happens if we consider $K_{Rice}=-10$ dB and $K_{Rice}=-100$ dB. For $K_{Rice}=0$ dB, the SNR needed to attain said target BER value lies between $1$ dB and $2$ dB apart with respect to the cases with $K_{Rice}=10$ dB and $K_{Rice}=-10$ dB. This is fully consistent with the differences in SNR we can appreciate in Fig. \ref{fig:Fig3MSE} for an MSE of $10^{-3}$, for example. We have also depicted the performance of the GEC-SR alternative in the case of the Rayleigh channel (i.e. $K_{Rice}=-100$ dB). As it may be seen, the performance is somewhat poorer than with our proposal for the same fading level. As a contrast with the behavior in the pure AWGN channel (see Fig. \ref{fig:Fig1BER}), the GEC-SR curve does not converge towards the curve of the EM-based system when the signal-to-noise ratio increases. This means that the GEC-SR algorithm is not so proficient when dealing with Rayleigh block fading, regardless of the AWGN level. It is to be noted that the developments in \cite{Yang19} consider a fast-fading model, and channel coefficients that change at each channel use. This could account for the specially bad performance under Rayleigh block fading.

\subsection{Average number of receiver iterations and steepest-descent steps}
\label{sec:iterations}

\begin{figure}[!ht]
\centering
\centerline{\includegraphics[width=\figurewidth]{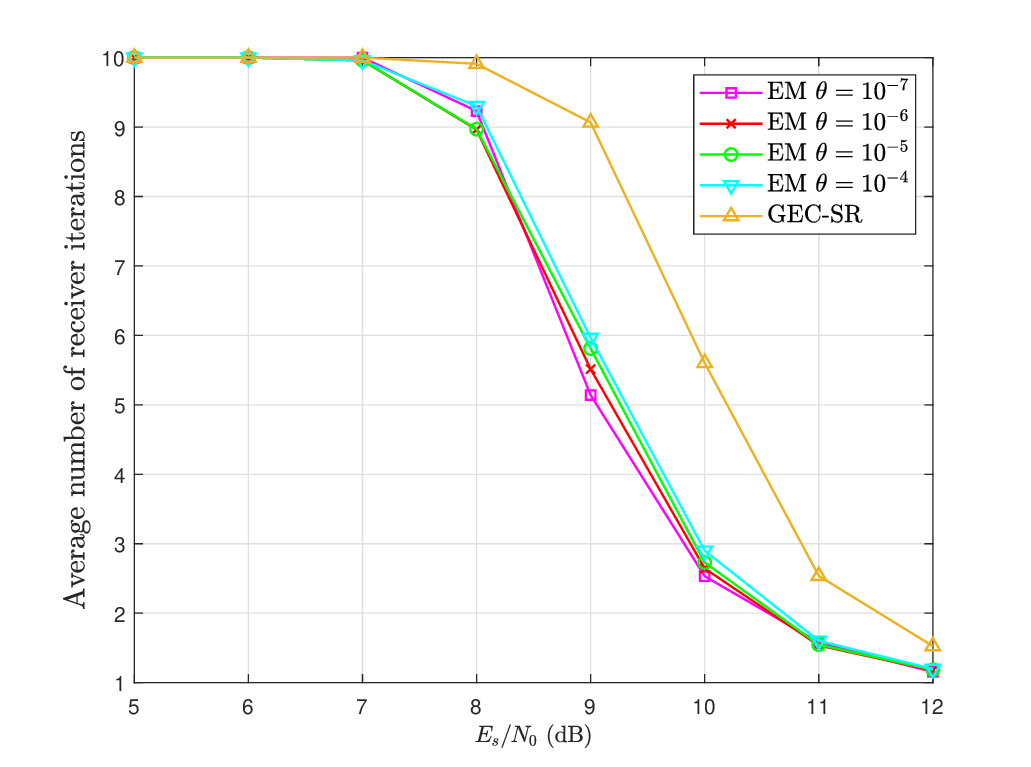}}
\caption{Average number of receiver iterations for several cases with different thresholds $\theta$ for the steepest-descent gradient algorithm. For comparison, the average number of receiver iterations for the GEC-SR alternative are also shown.}
\label{fig:Fig1Iter}
\end{figure}

\begin{figure}[!ht]
\centering
\centerline{\includegraphics[width=\figurewidth]{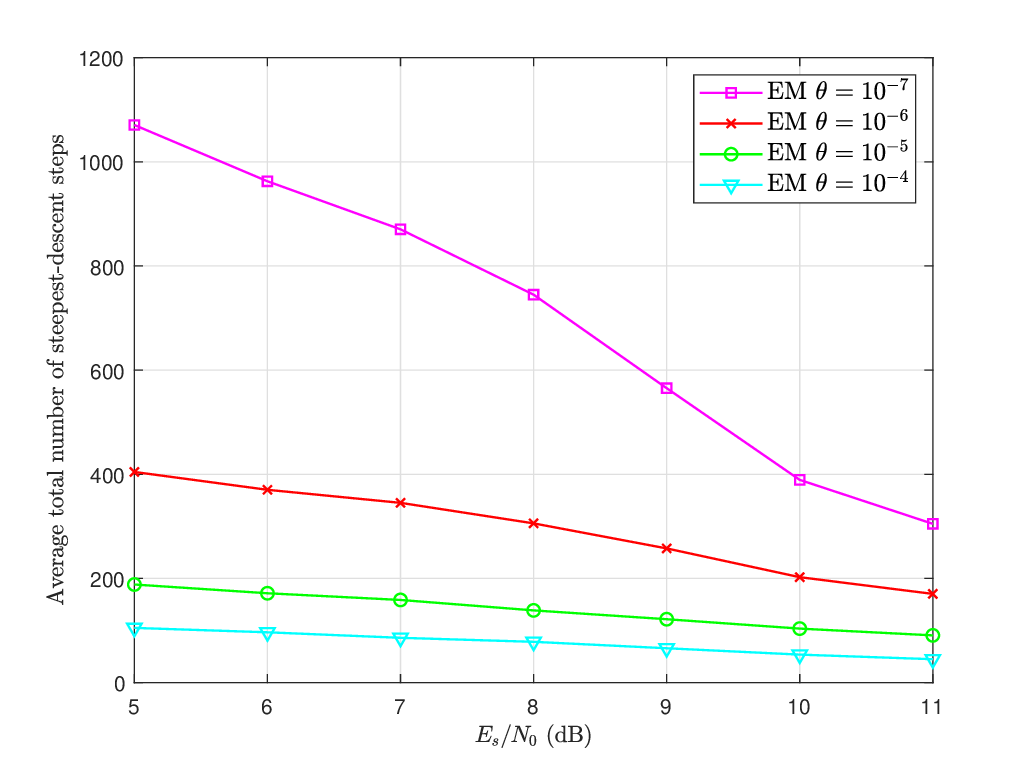}}
\caption{Average total number of steepest-descent gradient steps required as a function of threshold $\theta$ for the same cases considered in Fig. \ref{fig:Fig1Iter}, excepting the GEC-SR alternative.}
\label{fig:Fig2Iter}
\end{figure}

In Fig. \ref{fig:Fig1Iter}, we represent the average number of receiver iterations performed as a function of the SNR, in the nominal case (no fading, perfect CSI), for different thresholds $\theta$, compared with the GEC-SR alternative, and in Fig. \ref{fig:Fig2Iter} we depict the average total number of steepest-descent steps required. These figures allow understanding the computational requirements for the phase detection and correction process in each case, and may help in establishing important trade-offs. Fig. \ref{fig:Fig1Iter} shows how the threshold for the steepest-descent algorithm does not seem to affect much the average number of receiver iterations required to trigger the stopping rule as a function of the SNR. In general, we may distinguish three zones in the plots: for low SNR, the maximum number of receiver iterations (up to $10$) is reached because the stopping rule is not triggered; for an SNR range between roughly $7$ and $11$ dB, we have a steep descent in the number of iterations, a fact which is coherent with the activation of the stopping rule and the progressively better performance of the phase-detection algorithm and the decoding step; finally, for large SNR (from $12$ dB and on), we reach a floor where basically only one receiver iteration is performed, though from time to time up to two iterations would be performed in given cases and the average does not strictly fall to $1$ iteration. It is to be noted that, in the transition zone, there is a slight difference as a function of the threshold $\theta$ (for example, at $9$ dB): a case with a higher threshold (i.e., $\theta=10^{-4}$) requires an average number of iterations slightly higher than for lower values, and there is a consistent trend with decreasing $\theta$. This is logical, given that a lower threshold will provide a lower MSE (see Fig. \ref{fig:Fig1MSE}) and the stopping rule will be triggered earlier. Respecting the GEC-SR case, we can see there is a $1$-dB shift to the right, which means this algorithm requires a larger number of receiver iterations at equal SNR to attain a worse BER performance: for example, at $E_s/N_0 = 9$ dB, up to $3$ more iterations are required for GEC-SR, whereas the BER achieved is one order of magnitude higher, as seen in Fig. \ref{fig:Fig1BER}.

Respecting Fig. \ref{fig:Fig2Iter}, we can see a picture of how the steepest-descent algorithm performs throughout the detection process. We depict the {\it average total number} of steepest-descent steps for the same setup as in Fig. \ref{fig:Fig1Iter}, excepting the GEC-SR case, whose algorithm is completely different and is not directly comparable. We depict the total number of steepest-descent steps throughout all the receiver iterations, and it is an average respecting the number of frames processed. As it may be seen, and in contrast with Fig. \ref{fig:Fig1Iter}, there is always a non vanishing difference among the different thresholds, since, for a lower $\theta$, the algorithm will perform more steepest-descent steps regardless of the SNR level. In any case, the differences are higher for low SNR values (where a large amount of receiver iterations have to be performed), while, after the descent from $7$ to $11$ dB (which also reflects the trend identified in Fig. \ref{fig:Fig1Iter}), each case seems to stabilize around a value which decreases with higher values of $\theta$, as expected.

Though not shown, the BER results are basically the same for $\theta=10^{-7}$ and $\theta=10^{-6}$, while they slightly degrade for $\theta=10^{-5}$ and $\theta=10^{-4}$. This fact, together with the information provided by Figs. \ref{fig:Fig1Iter} and \ref{fig:Fig2Iter}, helps to identify as a convenient trade-off a threshold value of $\theta=10^{-6}$: it keeps the number of steps required as low as possible without noticeable BER degradation. It is worth noting that, for $\theta=10^{-6}$, the MSE is slightly worse with respect to the $\theta=10^{-7}$ case: this seeming contradiction probably arises from the fact that the channel decoder is a highly nonlinear, on-off device, which only works if its input has a quality exceeding a certain threshold, and this is not completely captured by the MSE. After such threshold is reached, a further improvement in the decoder input quality (as may be obtained by passing from $\theta=10^{-6}$ to $\theta=10^{-7}$) does not improve the BER anymore.

In Table \ref{table:IterEM}, we include the average number of receiver iterations, the average total number of steepest-descent steps and the SNR required for a BER of $10^{-4}$ in the case of other system setups, where there is either non-perfect CSI or a significant amount of fading (in all  cases, $\theta=10^{-6}$). We have included for comparison the figures for the $K_{Rice} = 100$ dB case with $\theta=10^{-6}$. As it can be expected, the SNR required grows as the detection or channel conditions get worse (i.e., lower $E_C/E_s$ or lower $K_{Rice}$), reflecting what we can see in the previous figures. The number of receiver iterations lies always between $4$ and $5$, reflecting the fact that the genie-aided decoding stopping criterion is triggered at a similar point for a given BER, regardless of other conditions. In the case of the average total number of steepest-descent steps, we have another picture. For the channel with Rician fading and perfect CSI (cases $(f)$ through $(i)$), we have practically the same values regardless of the level of fading, with a very slight increase as $K_{Rice}$ evolves towards the pure Rayleigh case.

If we focus on the non-fading imperfect CSI setup results (cases $(b)$ to $(e)$), we see that the trend is descending with lower values of $E_C/E_s$. This means that the number of steps required at a given BER decreases as the CSI estimation gets worse. The reason for this probably stands in the different behavior of the objective function \eqref{eq:Estep1_approx} with lower $E_C/E_s$ at the corresponding SNR. A higher amount of channel estimation noise is likely to smooth $h^{(l)}(\mathbf{\Phi})$, by filling the canyons and flattening the hills, thus yielding a faster convergence of the steepest-descent algorithm.

\begin{table*}[!ht]
    \centering
    \caption{Required $E_s/N_0$, average number of receiver iterations and average total number of steepest-descent steps at around $10^{-4}$ BER for the EM-based detector and the cases depicted in Figs. \ref{fig:Fig3BER} and \ref{fig:Fig4BER}. Case $(a)$: perfect CSI and $K_{Rice}=100$ dB. Cases $(b)-(e)$: no fading and $\frac{E_C}{E_s}=20$, $15$, $10$ and $5$ dB, respectively. Cases $(f)-(i)$: perfect CSI and $K_{Rice}=10$, $0$, $-10$, $-100$ dB, respectively.}
    \scriptsize
    \begin{tabular}{l|c|cccc|cccc}
    \hline
        ~ & $(a)$ & $(b)$ & $(c)$ & $(d)$ & $(e)$ & $(f)$ & $(g)$ & $(h)$ & $(i)$ \\
        \hline
        $E_s/N_0$ dB & $9.18$ & $10.55$ & $12.47$ & $15.67$ & $20.01$ & $9.44$ & $10.44$ & $11.85$ & $12.26$ \\
        Av. RX iterations & $4.87$ & $4.61$ & $4.54$ & $4.55$ & $4.61$ & $4.72$ & $4.74$ & $4.78$ & $4.51$ \\
        Av. total no. SD steps & $246.52$ & $218.71$ & $193.06$ & $153.97$ & $15.48$ & $243.22$ & $249.87$ & $250.44$ & $250.17$ \\
    \hline
    \end{tabular}
    \label{table:IterEM}
\end{table*}

\subsection{Algorithm complexity and latency}
\label{complexity}

The iterative receiver described in this paper has its core in the steepest-descent algorithm, used to solve the optimization problem in \eqref{eq:Phihat}, in order to perform the phase estimation. In the previous subsection, we have shown the average number of steepest-descent steps needed to trigger a convergence condition, as a function of threshold $\theta$, in various settings. 

In each step, for every time $n$, \eqref{eq:deri1} must be computed for each PN atomic process at the transmitter, \eqref{eq:deri2} for each process at the receiver.  Moreover, additional computation is related to the PN a-priori distribution (see \eqref{eq:ap}). Finally, the step size has to be optimized according to~\eqref{eq:BB} (except in the first iteration, where backtracking line search is performed). In Table~\ref{table:OpNum}, we show the number of operations that have to be performed by the proposed algorithm per steepest-descent step per time slot. To keep things simple, we have considered~\eqref{eq:BB} for the step size derivation.  In the same table,  we show the  figures for the GEC-SR phase detector, for which the number of operations per iteration per time slot is considered. In particular, we show in Table~\ref{table:OpNum} the number of sums, products, divisions, and LUT accesses, where the latter correspond to the implementation of functions such as exponentials, logarithms, sines and cosines, and so on.

\begin{table*}[!ht]
    \centering
    \caption{Number of operations per step/iteration per time slot for the steepest-descent phase detector (first column) and the GEC-SR (second column). In the first column, ~\eqref{eq:BB} is taken into account for step size computation.}
    \scriptsize
    \begin{tabular}{|l|c|c|}
    \hline
        Operation type & SD & GEC-SR \\ 
        \hline
        Sums & $4 N_t^2 + 8 N_r N_t + 12 N_{o,r} + 11 N_{o,t}$ & $23 N_r + 24 N_t + 6 N_t^2 + 8 N_r N_t + 11 M N_t + N_{o,t}$  \\
        Products & $5 N_t (N_t-1) + 10 N_r N_t + 7 (N_{o,r}+N_{o,t})$ & $53 N_r + 38 N_t + 12 N_t^2 + 13 N_r N_t + 7 M N_t$ \\
        Divisions & $N_{o,t} + N_{o,r}$ & $5 N_r + 5 N_t + 2 M N_t + N_{o,t} + N_{o,r} + 17$  \\ 
        LUT accesses & $ N_t (N_t-1) + 2 N_r N_t$ & $6 N_r + 8 N_t + 4 M N_t $ \\
    \hline
    \end{tabular}
    \label{table:OpNum}
\end{table*}

For the simulated case, $N_{o,t} = 16$, $N_t = 32$,  $N_{o,r} = 4$, $N_r = 64$, $M= 64$. Table~\ref{table:OpNumTot} shows the average number of mega-operations (millions of operations) per time slot for the steepest-descent detector and the GEC-SR with $E_s/N_0 =10$ dB.  For this SNR, the steepest-descent detector performs an average of 74.9 steps per receiver iteration, and an average 2.7 receiver iterations.  At the same SNR, the GEC-SR, which pays a performance penalty, performs 10.4 detector iterations per receiver iteration, and an average 5.6 receiver iterations. From the table, it can be seen that the steepest-descent detector requires about twice the number of operations required by the GEC-SR, with the notable exception of divisions, which for the steepest-descent are less than $1/50$ of those performed by the GEC-SR. The increase in complexity is the price to pay for the performance improvement.

\begin{table}[!ht]
    \centering
    \caption{Average number of mega-operations per time slot for the  steepest-descent phase detector (first column) and the GEC-SR (second column). }
    \scriptsize
    \begin{tabular}{|l|c|c|}
    \hline
        Operation type & SD & GEC-SR \\ 
        \hline
        Sums & $\simeq 4.192 $ & $\simeq 2.765 $  \\
        Products & $ \simeq 5.179 $ & $\simeq 3.381 $ \\
        Divisions & $\simeq 0.004$ & $\simeq 0.270 $  \\ 
        LUT accesses & $ \simeq 1.030 $ & $\simeq 0.516 $ \\
    \hline
    \end{tabular}
    \label{table:OpNumTot}
\end{table}

This detector complexity must be added to that of the other receiver blocks, especially of the channel decoder, which is typically the most complex block in the receiver. While this computational burden is not irrelevant, it has to be reminded that it is performed at the BS, where the hardware is usually powerful enough to bear it. 

The proposed receiver requires a certain number of iterations, thus it may seem to pose some challenges regarding latency. However, considering Fig. \ref{fig:Fig1Iter}, we notice that it boils down to a noniterative receiver for a large enough SNR\footnote{While we have used a genie-aided stopping rule to stop receiver iterations, a real stopping rule for channel decoders, e.g., LDPC decoders, performs in a similar way, just requiring a few more iterations on the average.}. Therefore, our proposal could be conceived as a tool to reduce the operative SNR, with respect to a noniterative receiver, at the price of some latency increase. In addition, if we suppose that users have different channel SNRs, we can imagine that those of them with a larger SNR would be decoded faster than the others, thus configuring a dynamical trade-off between channel quality and experienced latency. 

\section{Conclusions}

In this article we have proposed an EM-based algorithm for PN estimation and correction for a massive MIMO setup, with a general number of antennas per oscillator. We have also considered Rician block fading and imperfect CSI at the receiver side. The PN estimation algorithm is able to compensate for each atomic PN process stemming from each oscillator (both at the transmitter and at the receiver side). We have also characterized the BCRB for this setup, so as to be able to compare performances in terms of the MSE attained.

The core of the algorithm comprises iterative reception, demodulation and decoding, as well as a steepest-descent optimization stage with dynamic adaptation of the step and a stopping criterion based on a threshold.  We have simulated a massive MIMO system with different channel conditions and different setup parameters. The MSE results have shown that the algorithm can approach the BCRB for low values of the steepest-descent  threshold, but with little or no gain in BER below a given value. The results obtained using a more realistic PN model based on a mask, instead of the worst-case Wiener PN model, show that the system is robust and consistent in this respect. The MSE and BER results obtained for the imperfect CSI case show how the CSI mismatch can degrade the performance significantly, and stresses the fact that a sufficiently good channel estimation is essential in a massive MIMO framework.  Rician block fading was also been shown to hinder the performance, but with far less impact than the case of imperfect CSI. We have also compared the BER results obtained with our proposed phase detector with the results obtained resorting to another phase detector from the literature (based on the so-called GEC-SR algorithm \cite{Yang19}), and we have shown how the EM-based alternative outperforms the GEC-SR-based system.

On the other hand, we have evaluated the number of receiver iterations and steepest-descent algorithm steps required for given channel situations and specific configuration parameters. We have shown that it is possible to establish interesting trade-offs among MSE, BER and number of iterations/steps. In fact, by limiting the number of steepest-descent algorithm steps with an appropriate choice of the threshold, it is still possible to attain the best BER performance despite slight degradations in MSE. All this shows that the proposed receiver can be proficient in counteracting the effects of PN in massive MIMO systems, with an affordable computational burden and a trade-off between latency and received SNR, as compared with the GEC-SR alternative. In any case, as shown through the results and data shown in Subsections \ref{sec:iterations} and \ref{complexity}, in a potential implementation it will be critical to pay detailed attention to the resulting latency and the resource requirements for both algorithms (the EM-based and the GEC-SR based ones), and how they scale for specially large massive MIMO setups, in order to get viable systems. Future work will be focused in expanding the study of the properties of the system, for example, providing insights on how different alternatives for the massive MIMO demodulator and the channel code may impact the PN estimation and compensation process.

\section*{Acknowledgment}

This work was partially supported by the EU under the Italian NRRP of NextGenerationEU partnership on ``Telecommunications of the Future'' (PE00000001 - program ``RESTART'')

\ifCLASSOPTIONcaptionsoff
  \newpage
\fi

\end{document}